\def\citen#1{\if@filesw \immediate\write \@auxout {\string\citation{#1}}\fi%
\@tempcntb\m@ne \let\@h@ld\relax \def\@citea{}%
\@for \@citeb:=#1\do {\@ifundefined {b@\@citeb}%
    {\@h@ld\@citea\@tempcntb\m@ne{\bf ?}%
    \@warning {Citation `\@citeb ' on page \thepage \space undefined}}%
    {\@tempcnta\@tempcntb \advance\@tempcnta\@ne
    \setbox\z@\hbox\bgroup\ifcat0\csname b@\@citeb \endcsname \relax
    \egroup \@tempcntb\number\csname b@\@citeb \endcsname \relax
    \else \egroup \@tempcntb\m@ne \fi \ifnum\@tempcnta=\@tempcntb
    \ifx\@h@ld\relax \edef \@h@ld{\@citea\csname b@\@citeb\endcsname}%
    \else \edef\@h@ld{\hbox{--}\penalty\@highpenalty
    \csname b@\@citeb\endcsname}\fi
    \else \@h@ld\@citea\csname b@\@citeb \endcsname \let\@h@ld\relax \fi}%
\def\@citea{,\penalty\@highpenalty\hskip.13em plus.13em minus.13em}}\@h@ld}
\def\@citex[#1]#2{\@cite{\citen{#2}}{#1}}%
\def\@cite#1#2{\leavevmode\unskip\ifnum\lastpenalty=\z@\penalty\@highpenalty\fi%
  \ [{\multiply\@highpenalty 3 #1%
  \if@tempswa,\penalty\@highpenalty\ #2\fi}]}   %
\makeatother % end of CITE.STY
 
\catcode`\@=12
 
\def\A             {{\rm A}}

\def\alg           {algebra}
\def\alphad        {\tild\alpha}

\def\auto          {automorphism}
\def\betad         {\tild\beta}
\def\bc            {boundary condition}
\def\Bc            {Boundary condition}
\def\be            {\begin{equation}}
\def\bearl         {\begin{array}{l}}
\def\bearll        {\begin{array}{ll}}
\def\bearlll       {\begin{array}{lll}}
\def\bet           {\tilde\beta}
\def\bfe           {{\bf1}}
\def\blob          {\raisebox{.22em}{\rule{.25em}{.25em}}}
\newcommand\Breve[1]{#1^\omego}
\def\cala          {\chir}
\def\calap         {\chir^G}
\def\calc          {{\cal C}}
\def\calh          {{\cal H}}
\def\cals          {{\cal S}}

\def\Ce            {C\Untw}
\def\cft           {conformal field theory}
\def\Cft           {Conformal field theory}
\def\cfts          {conformal field theories}

\def\chid          {\tild{\chii}}
\def\chii          {\raisebox{.15em}{$\chi$}}

\def\chie          {\chii\Untw}
\def\chio          {\chii^\omega}
\def\chiO          {\chii\orb}
\def\chir          {\mbox{$\mathfrak A$}}

\def\chiz          {\chii\Tw}

\def\clAb          {\mbox{$\calc(\chir^\omega)$}}
\def\clAm          {\mbox{$\calc_-(\chir^\omega)$}}
\def\clAp          {\mbox{$\calc_+(\chir^\omega)$}}
\def\class         {classification}
\def\CO            {C\orb}

\def\Con           {Conformal }

\def\csa           {Cartan subalgebra}
\def\Cz            {C\Tw}
\def\df            {\,{:=}\,}

\def\Dim           {{\rm dim}\,}
\def\dl            {\mathbb }
\def\dsty          {\displaystyle}
\def\dyd           {Dynkin diagram}
\def\ee            {\end{equation}}
\def\eE            {{\rm e}}
\def\eear          {\end{array}}
\def\eps           {\epsilon}
\def\epsd          {{\tild\epsilon}}
\def\epso          {\varepsilon_\omego}
\def\epss          {\epsilon^\ssss}
\def\epsv          {\varepsilon}
\def\eq            {\,{=}\,}
\newcommand\erf[1] {(\ref{#1})}
\newcommand\Erf[2] {(\ref{#1#2})}
\def\etas          {\eta^\ssss}
\def\facrr         {\kappa_\omego}
\def\findim        {finite-dimensional}
\newcommand\fraC[2]{{#1}/{#2}}
\newcommand\Frac[2]{\mbox{\large$\frac{#1}{#2}$}}
\def\fsi           {Fro\-be\-ni\-us\hy Schur indicator}
\def\fts           {field theories}
\def\furu          {fusion rule}
\def\futnot#1      {}
\def\futnote#1     {\footnote{~#1}\ }
\def\g             {{\mbox{$\hat{\mathfrak g}$}}}
\def\gb            {\mbox{$\mathfrak g$}}

\def\gs            {{\omega^\star}}
\def\gv            {{{\rm g}^{\sss\vee}}}

\def\Gz            {Generalized }

\newcommand\hsp[1] {\mbox{\hspace{#1 em}}}
\def\hy            {$\mbox{-\hspace{-.66 mm}-}$}
\def\id            {\mbox{\sl id}}
\def\Id            {\mbox{\footnotesize\sl id}}
\def\ihwm          {irreducible highest weight module}
\def\ii            {{\rm i}}
\def\iN            {\,{\in}\,}

\def\Infdim        {Infinite-dimensional }

\def\irrep         {irreducible representation}
\def\J             {{\rm J}}

\def\Js            {\J_{\rm s}}
\def\Jv            {\J_{\rm v}}
\def\kma           {Kac\hy Moody algebra}
\long\def\labl#1   {\label{#1}\ee}
\long\def\Labl#1#2 {\label{#1#2}\ee}
\def\lambdab       {{\lambda}}
\def\lambdad       {{\tild\lambda}}
\def\lambdadp      {{\lambdad^+_{\phantom I}}}

\def\lambdap       {{\lambda^+_{\phantom I}}}

\def\lie           {Lie algebra}
\def\llb           {\mbox{\large(}}
\def\Llb           {\mbox{\Large(}}
\def\lolo          {L_0\ots\bfe+\bfe\ots L_0-c/12}
\def\lrb           {\mbox{\large)}}
\def\Lrb           {\mbox{\Large)}}
\def\LV            {L\raisebox{.51em}{$\sss\!\vee$}}
\def\LVi           {\LV_{\!\!\!\Id}}
\def\lvo           {L}
\def\LVo           {\LV_{\!\!\!\omego}}
\def\Lw            {L\raisebox{.51em}{${\sss\!\vee}\scs*$}}
\def\Lwi           {\Lw_{\!\!\!\!\!\Id}}
\def\lwo           {{L^{\!*}_{\phantom|}\!}}
\def\Lwo           {\Lw_{\!\!\!\!\!\omego}}
\def\Ltimes        {{\ltimes}} 
\def\Mapsto        {\,{\mapsto}\,}
\def\modinv        {modular invarian}

\def\mud           {{\tild\mu}}

\def\MV            {\tild L\raisebox{.51em}{$\sss\!\vee$}}
\def\MVi           {\MV_{\!\!\!\Id}}
\def\mvo           {{\tild L}}
\def\MVo           {\MV_{\!\!\!\omego}}
\def\Mw            {\tild L\raisebox{.51em}{${\sss\!\vee}\scs*$}}
\def\Mwi           {\Mw_{\!\!\!\!\!\Id}}
\def\mwo           {{\tild L^{\!*}_{\phantom|}\!}}
\def\Mwo           {\Mw_{\!\!\!\!\!\omego}}
\newcommand\N[3]   {{\rm N}_{#1,#2}^{\;\ \ #3}}
\def\nE            {\,{\ne}\,}
\newcommand\Nel[3] {{\rm N}\Untw_{#1,#2,#3}}
\newcommand\Nl[3]  {{\rm N}_{#1,#2,#3}}
\def\nn            {$N\,{=}\,2$ }
\newcommand\No[3]  {{\rm N}_{#1,#2}^{\Orb\,\ \ \ \ \ \ \ \ \ #3}}
\newcommand\Nol[3] {{\rm N}\orb_{#1,#2,#3}}

\def\nxl           {\\{}&&&&&\\[-.81em]}
\def\nxL           {\\[.47em]\hline &&&&&\\[-.81em]}
\newcommand\nxt[1] {\\\raisebox{.12em}{\rule{.35em}{.35em}}\hsp{.6}#1}

\def\oc            {{\omega_{\rm c}}}
\def\omego         {{\omega_\circ}}
\def\omegv         {{\wob}}
\def\one           {\mbox{\small $1\!\!$}1}

\def\oplu          {\,{\oplus}\,}
\def\oplU          {{\oplus}}
\def\orb           {^\Orb}
\def\Orb           {{\cal O}}

\def\ots           {\raisebox{.07em}{$\sss\otimes$}}
\def\P             {P^{+}}
\def\parfu         {partition function}

\def\Pd            {\tild P^{\omego\sss+}}
\def\Po            {P^{\omego\sss+}}
\def\PPd           {\tild P^{\omego\sss++}}
\def\PPo           {P^{\omego\sss++}}
\def\psj           {\tilde\psi}
\def\pslz          {\mbox{PSL$(2{,}\zet)$}}

\def\rank          {{\rm rank}\,}
\def\rcft          {rational conformal field theory}
\def\reals         {{\dl R}}
\def\rep           {representation}
\def\resp          {respectively}

\def\Rhob          {\rho}
\def\Rhod          {\tild\Rhob}

\def\scs           {\scriptstyle}

\def\Se            {S\Untw}
\newcommand\sect[1]{\section{#1}\setcounter{equation}{0}}
\def\sigm          {{f_\omego}}
\def\sigmas        {{\sigma_{\!s}}}

\def\so            {\mathfrak{so}}

\def\SO            {S\orb}
\def\SOTOSO        {X\orb}
\def\sp            {s_\circ}
\def\sss           {\scriptscriptstyle}
\def\ssss          {{\scriptscriptstyle[s]}}

\def\su            {\mathfrak{su}}
\def\suco          {superconformal }

\newcommand\sumeq[1]{\sum_{\scs #1 \atop #1 = \gs #1}}
\newcommand\sumeqd[1]{\sum_{\scs \tild{#1}}}

\def\sLV           {L^{\!\vee}}
\def\sLVo          {\sLV_\omego}

\def\sMV           {\mvo^{\!\vee}}
\def\sMVo          {\sMV_\omego}

\def\syms          {sym\-me\-tries}
\def\Sz            {S\Tw}
\def\Tau           {{\cal T}^\omega}

\def\Tauo          {{\cal T}^\omego}

\def\Te            {T\Untw}
\def\Thetad        {\tild\Theta}
\def\tild          {\dot}
\def\Tilde         {\tilde}

\newcommand\tN[6]  {\Tilde{\rm N}_{(#1,#2),(#3,#4)}^{\ \ \ \ \ \ \ \ \ (#5,#6)}}

\def\TO            {T\orb}
\def\Tr            {{\rm Tr}}
\def\tS            {\Tilde S}
\def\tw            {^{\sss(2)}}
\def\Tw            {^{\sss(1)}}
\def\twodim        {two-dimensional}
\def\Tz            {T\Tw}
\def\untw          {^{\sss(1)}}
\def\Untw          {^{\sss(0)}}
\def\uone          {\mbox{$\mathfrak{u}(1)$}}
\def\vac           {\Omega}
\def\varp          {\tilde\omega}

\def\Vee           {{\sss\vee}}

\def\voa           {vertex operator algebra}

\def\w             {\hat w}
\def\W             {\hat W}
\def\wb            {w}
\def\Wb            {W}

\def\Wo            {{\W_{\!\omego}}}
\def\Wib           {{\Wb_{\!\Id}}}
\def\wob           {{\cal W}}
\def\Wob           {{\Wb_{\!\omego}}}
\def\wrt           {with respect to }
\def\wrtt          {with respect to the }
\def\wzwm          {WZW model}
\def\wzwt          {WZW theory}
\def\wzwts         {WZW theories}
\def\Xid           {\tild\Xi}
\def\zet           {{\dl Z}}
\def\Zeta          {\zeta_\omego}
\def\zetplus       {{\dl Z}_{>0}}
\def\zetpluso      {{\dl Z}_{\ge0}}
 
\newcounter{fnot}

\documentclass[12pt]{article} \usepackage{amssymb,amsfonts,latexsym}

\begin{document}
 
%%%%%%%%%%%%%%%%%%%%%%%%%%%%%%%%%%%%%%%%%%%%%%%%%%%%%%%%%%%%%%%%%%%%%%%

\begin{flushright}  {~} \\[-1cm] 
{\sf hep-th/9905038} \\[1mm] {\sf ETH-TH/99-11} \\[1 mm]
{\sf May 1999} \end{flushright}

\begin{center} \vskip 14mm
{\Large\bf SYMMETRY BREAKING BOUNDARY}\\[4mm]
{\Large\bf CONDITIONS AND WZW ORBIFOLDS}\\[22mm]
{\large Lothar Birke, \ J\"urgen Fuchs, \
Christoph Schweigert}\\[7mm] Institut f\"ur Theoretische Physik \\[2mm]
ETH H\"onggerberg \\[2mm] CH -- 8093~~Z\"urich
\end{center} \vskip 21mm
\begin{quote}{\bf Abstract}\\[1mm]
Symmetry breaking boundary conditions for WZW theories are discussed.
We derive explicit formulae for the reflection coefficients in the presence 
of boundary conditions that preserve only an orbifold subalgebra with 
respect to an involutive automorphism of the chiral algebra. The characters 
and modular transformations of the corresponding orbifold theories are 
computed. Both inner and outer automorphisms are treated.
\end{quote}
\newpage

%%%%%%%%%%%%%%%%%%%%%%%%%%%%%%%%%%%%%%%%%%%%%%%%%%%%%%%%%%%%%%%%%%%%%%%

\sect{Introduction}

Our understanding of conformally invariant boundary conditions for \twodim\ 
\cfts\ and of their classification has recently improved enormously. Such 
boundary conditions allow to study string perturbation theory in the background
of certain solitonic solutions, so-called D-branes. They also possess 
applications in statistical mechanics, e.g.\ in the description of impurities 
and percolation problems. 

For non-trivial \cfts, i.e.\ for string backgrounds that are not flat,
conformal field theory techniques have been applied successfully in
those situations where only finitely many primary fields occur. More precisely,
these are cases where not only the bulk theory is rational, but also the 
part $\calap$
of the bulk symmetry $\cala$ that is not broken by the boundary conditions
is still the chiral algebra of a rational theory. The situation is
particularly manageable when $\calap$ is actually an orbifold subalgebra,
i.e.\ a subalgebra of $\cala$ that is left pointwise fixed by some group
$G$ of automorphisms. For the case when $G$ is a finite 
abelian group, an explicit description and classification
of such boundary conditions has been established
in \cite{fuSc10,fuSc11}. The crucial ingredients in those investigations
are the representations of the modular group, both the one that is associated 
to the chiral \cft\ based on $\cala$ and the one associated to the
orbifold chiral algebra $\calap$.

The purpose of the present note is to exploit these novel results in the 
special case when $\cala$ is the chiral algebra of a WZW theory. In this case
the theory affine Lie algebras provides a powerful tool to compute the
relevant modular matrices. There are, however, also other reasons to study 
this specific class of models. It has been conjectured \cite{fuSc6} that the
structure constants of the classifying algebra for the boundary conditions
of a given automorphism type are given by the traces of the action of this
automorphism on the spaces of chiral blocks. In the case of \wzwts, 
techniques are available (see e.g.\ \cite{tsuy,beau})
that allow to test this conjecture. Moreover, WZW theories correspond to 
strings propagating on group manifolds; thus they constitute the most directly 
accessible non-trivial generalization of strings propagating in a flat 
background. Boundary conditions of WZW theories therefore present a convenient 
testing ground for studying conjectures about the correspondence of geometric 
and algebraic formulations of boundary conditions. While in the case of flat 
backgrounds one deals with the familiar Neumann and Dirichlet \bc s of free 
bosons (possibly supplemented with a background field strength), already in 
this modestly generalized situation the geometric interpretation of 
boundary conditions, in particular of those which break bulk symmetries, 
still remains to be clarified.

As already mentioned, the analysis of the boundary conditions of our interest
requires in particular a rather detailed knowledge of the modular matrices of the
orbifold theory. As a consequence, we first have to establish these data,
which in itself constitutes a non-trivial result. In the technically
more amenable case where the orbifold group $G$ consists of inner automorphisms,
such orbifolds have been studied earlier \cite{kaTo2}. In the present
paper, we will also determine the modular matrices for orbifolds by outer
automorphisms. As it turns out, the twisted sector of such orbifold theories
is provided by integrable \irrep s of {\em twisted\/} affine \lie s.
As a simplification, here we restrict ourselves to the case of orbifolds by 
involutions, i.e.\ to $G\eq\zet_2$.
But as a matter of fact our results on WZW orbifolds are already sufficient
to extract also the reflection coefficients for more general symmetry
breaking boundary conditions.

Once the modular matrices of the orbifold theory have been obtained,
the results of \cite{fuSc10,fuSc11} allow to read off the corresponding
symmetry breaking boundary conditions rather directly. In fact, we 
learn even more; by applying T-duality both in the bulk and to the
\bc s we can also describe
boundary conditions that preserve all symmetries of the bulk theory, but
for a theory with non-trivial modular invariant in the bulk.
Our results cover in particular the case of orbifolds by
the charge conjugation automorphism. Thus, by T-duality,
we are able to describe the boundary conditions that arise for the true diagonal
torus \parfu. (Typically, in the literature boundary conditions are 
considered for the charge conjugation modular invariant 
\cite{card9,fuSc6,fuSc11}; see, however, \cite{prss2,fuSc5}.)

The paper is organized as follows. We start in section 2 by summarizing
the general structure of $\zet_2$-orbifolds, with emphasis on the building 
blocks of the orbifold characters and their modular transformations.
Section 3 contains the technical core of the paper. We establish the 
techniques for dealing with arbitrary involutive automorphisms of the 
chiral \alg\ of a \wzwt\ that leave the Virasoro element fixed,
and compute the characters and modular matrices for the orbifold
of the \wzwt\ by such an automorphism. In section 4 these
results are combined with the information from \cite{fuSc11} so as to 
determine the symmetry breaking boundary conditions of the \wzwt.
Finally, an appendix collects some properties of twisted Theta functions.

\sect{$\zet_2$-orbifolds}

\subsection{Elementary consistency conditions}

We study the orbifold theory of an arbitrary rational \cft\ by the
$\zet_2$-group that is generated by an order-two \auto\ 
  \be  \omega:\quad \chir\to\chir \,, \qquad \omega^2\eq\id \,,  \Labl2z
of the chiral \alg\ \chir\ of the original theory. The chiral algebra
of the orbifold theory is by definition the subalgebra $\chir^{\zet_2}$ of 
\chir\ that is left pointwise fixed by $\omega$. 
To give rise to a consistent \cft, this orbifold subalgebra must again
possess a Virasoro element. Conformal invariance of the boundary conditions
requires in addition that the Virasoro elements of \chir\ and $\chir^{\zet_2}$
actually coincide; thus we require that the automorphism $\omega$ leaves 
the Virasoro element fixed.

For any arbitrary \auto\ $\varp$ of \chir\ and any \chir-\rep\ $R$ also 
$R\,{\circ}\, \varp$ is a \rep\ of \chir. As a consequence, associated 
with $\varp$ there comes a bijection $\varp^\star$ between \irrep s of \chir\
that is defined by $R_\lambda{\circ}\,\varp\,{\cong}\,
R_{\varp^\star\lambda}$. Since the orbifold chiral algebra $\chir\orb$ only 
contains elements that are invariant under $\omega$, there are no observables
in the orbifold theory that could distinguish between $R_\lambda$ and 
$R_{\varp^\star\lambda}$. These representations, even when they are inequivalent
representations of the original chiral algebra \chir, will give equivalent
representations of $\chir\orb$. This identification is, however, not the only
effect. Rather, when $R_{\varp^\star\lambda}\,{\cong}\,R_\lambda$, then the 
automorphism $\varp$ is implemented by an \auto\ of the \chir-module
$R_\lambda$, and the invariant subspaces under this map
constitute sub{\em modules\/} for $\chir\orb$. As a consequence, one is faced
with the task to split these \chir-modules into submodules over $\chir\orb$.
This way we arrive at a certain set of irreducible $\chir\orb$-modules, but 
typically we do not get all of the irreducible $\chir\orb$-modules. The ones 
which can be obtained by identifying
and splitting \chir-modules constitute the untwisted sector of the orbifold
theory, while all other \rep s of $\chir\orb$ are said to be in some
twisted sector. We remark that by definition the untwisted sector is closed
under operator products, but it is not closed under modular transformations.
This very fact will enable us to determine the twisted sector.

Let us now specialize to the $\zet_2$ orbifold obtained for the \auto\ 
\Erf2z. We first consider the untwisted sector.
There are two types of fields, which are distinguished by the action
of the orbifold group on the fields of the original theory they come
from. More specifically, they differ in the size of the {\em stabilizer\/}
subgroup
  \be  S_\lambda := \{ \omega\iN\zet_2 \,|\, \omega^\star\lambda\eq\lambda
  \} \,.  \ee
When $S_\mu\eq\zet_2$, then we call the primary field $\mu$ {\em symmetric\/},
while for $S_\mu\eq\{\id\}$, we call $\mu$ {\em non-symmetric\/}.\,%
 \futnote{In the case of permutation orbifolds, such fields have been called 
\cite{bohs} {\em diagonal\/} and {\em off-diagonal\/}, \resp.}
For non-symmetric $\mu$, the two fields on the $\zet_2$-orbit
$\{\mu,\gs\mu\}$ are isomorphic modules of the orbifold chiral algebra.
As a consequence,
each non-symmetric orbit gives rise to a single primary field in the orbifold;
we choose (arbitrarily) a representative $\lambda$ for each such length-two 
orbit $\{\lambda,\gs\lambda\}$ and label the corresponding orbifold field as 
$(\lambda,0,0)$; its character is simply
  \be  \chiO_{(\lambda,0,0)}(\tau) = \chii_\lambda^{}(\tau) \,.  \Labl tw
In contrast, each symmetric field $\lambda$ of the original theory, 
satisfying $\gs\lambda\eq\lambda$, gets
split, thus giving rise to two distinct fields in the untwisted sector of 
the orbifold; we label them as $(\lambda,\psi,0)$ with $\psi\iN\{\pm1\}$.
Their characters are to be obtained by a suitable projection, and
accordingly can be written as
  \be  \chiO_{(\lambda,\psi,0)}(\tau) = \Frac12\, \llb \chii_\lambda(\tau)
  + \psi\, \eta_\lambda^{-1}\, \chie_\lambda(2\tau) \lrb  \labl{utw}
with certain phases $\eta_\lambda$. At this point is just a definition 
of the expression $\eta_\lambda^{-1} \chie_\lambda$ that is hereby 
introduced for every symmetric field; 
this definition implies in particular that
  \be \chie_\lambda(\tau{+}2) = (\Te_\lambda)^2\, \chie_\lambda(\tau)
  \qquad{\rm with}\qquad (\Te_\lambda)^2 = \TO_{(\lambda,\psi,0)}
  = T_\lambda^{} \,,  \labl{fact}
where $T$ denotes the T-matrix of the original theory.
Also, the explicit introduction of the phases $\eta_\lambda$ is not
really necessary, but this will prove to be convenient later on;
we take the convention, though, to ascribe the value $\eta_\vac\eq1$ to 
the phase for the vacuum primary field $\vac$.

We can also immediately present some S-matrix elements
of the orbifold theory, namely those for the S-transformation of the
characters \Erf tw coming from non-symmetric fields; they are expressible
through the S-matrix $S$ of the original theory as
  \be \bearl
  \SO_{(\lambda,0,0),(\mu,0,0)} = S_{\lambda,\mu} + S_{\lambda,\gs\mu} \,,
  \\{}\\[-.8em]
  \SO_{(\lambda,0,0),(\mu,\psi,0)} = S_{\lambda,\mu}  \,,
  \\{}\\[-.8em]
  \SO_{(\lambda,0,0),(\mud,\psi,1)} = 0  \,.
  \end{array} \Labl1s
Here in the last line we introduced the notation $(\lambdad,\psi,1)$ for
the fields in the twisted sector of the orbifold that come from
the symmetric field $\lambda$; each symmetric field gives rise to
two such fields, distinguished by the value of $\psi\iN\{\pm1\}$, while
there are no twisted fields coming from non-symmetric fields $\lambda$.

In order for the result in the second line of \Erf1s to be well-defined,
i.e.\ independent on whether one takes the representative $\lambda$ or
$\gs\lambda$ of a non-symmetric orbit, we need $S_{\lambda,\mu}\eq 
S_{\gs\lambda,\mu}$ for every symmetric $\mu$; this is a condition that 
must be satisfied for every consistent orbifold action. Further, to
have a consistent orbifold theory, $\SO$ must be symmetric; this implies that
$S_{\lambda,\mu}{+}S_{\lambda,\gs\mu}\eq S_{\mu,\lambda}{+}S_{\mu,\gs\lambda}$,
which together with the symmetry of the original S-matrix $S$ implies that 
the previous requirement generalizes to
  \be  S_{\lambda,\gs\mu} = S_{\gs\lambda,\mu} \labl{fund}
for every $\mu$. Thus for every consistent $\zet_2$-orbifold
action, the induced map on the primaries must have the property that
\erf{fund} holds for all $\lambda$ and $\mu$.
Moreover, in order for the orbifold to furnish a \cft\ the automorphism 
$\omega$ of the chiral \alg\ must keep the Virasoro algebra fixed, which
in turn implies that it maps the vacuum primary field $\vac$ to itself,
$\gs\vac\eq\vac$. 
These elementary results for the orbifold tell us in particular that 
  \be  \sum_\kappa \Frac{S_{\gs\lambda,\kappa}^{}S_{\gs\mu,\kappa}^{}
  S_{\gs\nu,\kappa}^*} {S_{\gs\vac,\kappa}}
  = \sum_\kappa \Frac{S_{\lambda,\gs\kappa}^{}S_{\mu,,\gs\kappa}^{}
  S_{\nu,\gs\kappa}^*} {S_{\vac,\gs\kappa}} \,,  \ee
which by the Verlinde formula means that
the fusion rules of the original theory satisfy
  \be \N{\gs\lambda}{\gs\mu}{\ \ \ \gs\nu} = \N\lambda\mu\nu \,.  \ee
Thus the map $\gs$ constitutes an {\em automorphism of the fusion rules\/}.
One should, however, be aware of the fact that different automorphisms 
of the chiral algebra can give rise to one and the same automorphism of the 
fusion rules; to give an example, we will see that every inner automorphism 
of a \wzwt\ yields the identity map on the fusion rule \alg.

\subsection{Characters in the twisted sector}

To be able to analyze also the
characters in the twisted sector, additional structure is needed. 
The characters in the twisted sector can be obtained by performing
an S-transformation $\tau\Mapsto{-}1/\tau$ on the functions $\chie$. 
   The result can be written as a linear combination of 
   character-like quantities. In all cases known to us 
   each of those is, up to an over-all power of $q\eq\exp(2\pi\ii\tau)$, a
   power series in $q^{1/2}$, albeit not in $q$. 
Each such series arises from the sum of characters of two primary fields in 
the twisted sector, and the two characters will be obtained separately as 
the eigenfunctions of the T-operation $\tau\Mapsto\tau{+}1$. 

Therefore we demand that for every
symmetric field $\lambda$ there is another function
  \be  \chiz_\lambdad(\tau) \,,  \ee
such that the following relations are obeyed.
\nxt
The labels $\lambdad$ are in one-to-one correspondence with the labels
$\lambda$. (Generically this one-to-one correspondence is, however, not
canonical. In particular, in general one cannot dispense of using
two different kinds of symbols for the labels of the functions $\chie$
and $\chiz$.)
\nxt
The function $\chiz_\lambdad$ has non-negative integral coefficients 
in its expansion in powers of $q\eq\exp(2\pi\ii\tau)$.
\nxt
The functions are eigenfunctions under T, i.e.\ we have
  \be  \chiz_\lambdad(\tau{+}1)
  = \Tz_\lambdad\, \chiz_\lambdad(\tau)  \labl{a1}
for some suitable phases $\Tz_\lambdad$.
\nxt
The S-operation connects the functions $\chie_\lambda$ and $\chiz_\mud$.
More precisely, there are unitary matrices $\Se$ and $\Sz$ such that
  \be \chie_\lambda(-\Frac1\tau) = \sum_\mud \Se_{\lambda,\mud}\,
  \chiz_\mud(\tau) \Labl a2
and 
  \be \chiz_\lambdad(-\Frac1\tau) = \sumeq\mu
  \Sz_{\lambdad,\mu}\, \chie_\mu(\tau) \,.  \Labl a3
This structure is present in all examples that are known to us. 
Note that even when the labels $\lambda$ and $\lambdad$ take their values
in the same set, the functions $\chie$ and $\chiz$ are definitely not required 
to coincide. Thus even in this special situation there is no reason
to require that the matrices $\Se$ and $\Sz$ are symmetric,
though this will be the case in specific examples.

Combining the conditions \Erf a1 and \Erf a2,
we find the characters in the twisted sector as
  \be  \chiO_{(\lambdad,\psi,1)}(\tau)
  = \Frac12\, \llb \chiz_\lambdad(\Frac\tau2) + \psi\,
  (\Tz_\lambdad)^{-1/2}_{}\, \chiz_\lambdad(\Frac{\tau+1}2) \lrb \,.  \Labl Tw
They transform under T as
  \be \chiO_{(\lambdad,\psi,1)}(\tau{+}1) = \psi\, (\Tz_\lambdad)^{1/2}_{}\,
  \chii_{(\lambdad,\psi,1)}(\tau) \,, \ee
which determines the conformal weights (modulo integers) in the twisted sector,
  \be  \TO_{(\lambdad,\psi,1)} = \psi\, (\Tz_\lambdad)^{1/2}_{}\,.  \ee
Note that the square root $(\Tz_\lambdad)^{1/2}$ is only defined up to a sign;
it is to be understood as follows. For each value of $\lambdad$ we
make an arbitrary choice of the square root, and keep
this choice fixed once and forever. A different choice would lead to the 
opposite assignment of the label $\psi$. Once this choice has been made,
fields with the same $\lambdad$, but different values of $\psi$, can be 
distinguished
unambiguously, because their conformal weights differ by $1/2\bmod\zet$.

We are now in a position to determine the rest of the S-matrix
elements in the untwisted sector. We obtain
  \be \bearl
  \SO_{(\lambda,\psi,0),(\mu,\psi',0)} = \Frac12\, S_{\lambda,\mu} \,,
  \\{}\\[-.8em]
  \SO_{(\lambda,\psi,0),(\mud,\psi',1)} =
  \Frac12\, \psi\eta_\lambda^{-1}\, \Se_{\lambda,\mud} \,,
  \\{}\\[-.8em]
  \SO_{(\lambda,\psi,0),(\mu,0,0)} = \Frac12\, (S_{\lambda,\mu} +
  S_{\lambda,\gs\mu}) = S_{\lambda,\mu} \,.
  \end{array}\ee
(The first and third lines do not rely on our assumptions about the 
functions $\chiz$.)

For the modular S-transformation of the characters \Erf Tw we find
  \be  \chii_{(\lambdad,\psi,1)}(-\Frac1\tau) 
  = \Frac12 \dsty\sum_\mu \sum_{\psi'=\pm1} \psi' \llb
  \eta_\mu^{}\, \Sz_{\lambdad,\mu}\, \chii_{(\mu,\psi',0)}(\tau)
  + \psi\, P_{\lambdad,\mud}\, \chii_{(\mud,\psi',1)}(\tau) \lrb
  \,,  \Labl,1
where $P$ is the matrix
  \be  P := (\Tz)^{1/2}_{} \Sz (\Te)^2_{} \Se (\Tz)^{1/2}_{}  \Labl0P
(the symbol $P$ is chosen in accordance with the notation for a similar
matrix that appeared in a different context in \cite{prss}). By the 
unitarity of the various matrices appearing here, $P$ is unitary as well.
The precise form of the matrix $P$ depends on our choice of
square roots in the definition of $\TO$.
But the result for the matrix $\SO$ is independent of our choice of
square roots, because for a different 
choice we also must flip the labelling of characters, so that
the expression $\psi\psi'P_{\lambdad,\mud}$ remains unchanged.
{}From \Erf,1 we read off the remaining elements of the modular matrix $\SO$
of the orbifold:
  \be \bearl
  \SO_{(\lambdad,\psi,1),(\mu,\psi',0)}
  = \Frac12\, \eta_\mu^{} \psi'\,\Sz_{\lambdad,\mu} \,,
  \\{}\\[-.8em]
  \SO_{(\lambdad,\psi,1),(\mud,\psi',1)} = \Frac12\,\psi\psi'\,P_{\lambdad,\mud}
  \,, \\{}\\[-.8em]
  \SO_{(\lambdad,\psi,1),(\mu,0,0)} = 0
  \,. \end{array}\ee
Thus in particular we have managed to express, based on only very general 
assumptions,
the matrix elements $\SO_{(\lambdad,\psi,1),(\mud,\psi',1)}$ in the twisted
sector in terms of other $\SO$-elements and data of the original theory.

\subsection{Consistency conditions}

In order for the orbifold construction to furnish a consistent \cft, the 
matrices $\SO$ and $\TO$ must possess the following properties:
  \be  \bearlll
  \blob& {\rm symmetry}\,{:}    & {(\SO)}^{\rm t} = \SO \,,  \\{}\\[-.7em]
  \blob& {\rm unitarity}\,{:}   & {(\SO)}^{-1} = {(\SO)}^* \,,  \\{}\\[-.7em]
  \blob& {\rm conjugation}\,{:} & \CO := (\SO)^2 \;\ \mbox{is an order-two
                                  permutation} \,, \\{}\\[-.7em]
  \blob& & {(\SO\TO)}^3 = {(\SO)}^2 \,.
  \eear \Labl ST
In addition, when inserting $\SO$ into the Verlinde formula, we must obtain
non-negative integral fusion coefficients.

Let us discuss the implications of the requirements \Erf ST in some detail.

\subsubsection{Symmetry of $\SO$}

To investigate the symmetry property of the matrix $\SO$, we compare its
entry $\SO_{(\lambda,\psi,0),(\mud,\psi',1)}$ to 
$\SO_{(\mud,\psi',1),(\lambda,\psi,0)}$. 
One sees that a necessary condition for symmetry of $\SO$ is that
  \be  \Sz_{\mud,\lambda} = \eta_\lambda^{-2} \Se_{\lambda,\mud} \,.  \labl{scon}
This relation is compatible with the unitarity of the matrices because
the numbers $\eta_\lambda$ are phases. It also allows us to rewrite $P$ as
  \be  P = (\Tz)^{1/2}_{} (\Se)^{\rm t}_{} (\eta^{-1}\Te)^2_{} \Se (\Tz)^{1/2}_{}
  \,.  \labl{sym}
This is manifestly symmetric, which implies that \erf{scon} is also sufficient
for $\SO$ to be symmetric.

\subsubsection{Unitarity of $\SO$ and charge conjugation}

The conditions that ensure that $\SO$ is unitary
and that the charge conjugation $\CO\eq(\SO)^2$ is a permutation of order
two are best studied together. For non-symmetric $\lambda$, straightforward 
calculation using the 
unitarity of $S$ shows that $(\SO{(\SO)}^*)_{(\lambda,0,0),(\lambda',0,0)}
\eq\delta_{(\lambda,0,0),(\lambda',0,0)}$ and
$(\SO (\SO)^*)_{(\lambda,0,0),(\lambda',\psi,g)}\eq0$ are satisfied
automatically. To determine the conjugation for non-symmetric $\lambda$,
we just use that $\gs$ is a fusion rule \auto\ and hence
commutes with charge conjugation; it follows that
charge-conjugate $\gs$-orbits give charge-conjugate fields
in the orbifold theory. (Thus when $\gs$ is itself charge conjugation,
all fields $(\mu,0,0)$ of the orbifold theory are selfconjugate.)

For symmetric fields, the desired result
$(\SO(\SO)^*)_{(\lambda,\psi,0),(\lambda',\psi',0)} 
\eq\delta^{}_{(\lambda,\psi,0),(\lambda',\psi',0)}$
is obtained if and only if $\Se$ and $\Sz$ are related as
  \be \sum_\mud \Se_{\lambda,\mud}\, (\Sz_{\mud,\lambda'})^*_{}
  = \eta_\lambda^2\, \delta^{}_{\lambda,\lambda'} \,,  \ee
or equivalently,
  \be  \sumeq\mu \Sz_{\lambdad,\mu}\,\eta_\mu^2\,(\Se_{\mu,\lambdad'})^*
  _{} = \delta_{\lambdad,\lambdad'}^{} \,. \ee
As for charge conjugation, we obtain
  \be  \CO_{(\lambda,\psi,0),(\lambda',\psi',0)} 
  = \Frac12\, ( C_{\lambda,\lambda'}^{}
  + \psi\psi'\, \Ce_{\lambda,\lambda'} )  \ee
with
  \be  \Ce_{\lambda,\lambda'} := \eta_\lambda^{-1} \eta_{\lambda'}^{-1}
  \sum_\mud \Se_{\lambda,\mud} \Se_{\lambda',\mud}
  = \eta_\lambda^{-1} \eta_{\lambda'}^{}\,(\Se\Sz)_{\lambda,\lambda'}
  \,.  \ee
Thus charge conjugation in the orbifold is consistent only if it is 
possible to choose sign factors 
  \be  \eps_\lambda \in \{\pm1\}  \ee
in such a manner that 
  \be  \Ce_{\lambda,\lambda'} = \eps_\lambda^{} C_{\lambda,\lambda'}^{}
  \,.  \ee

For the twisted sector, we find that
$(\SO{(\SO)}^*)_{(\lambdad,\psi,1),(\lambdad',\psi',1)}
\eq\delta_{(\lambdad,\psi,1),(\lambdad',\psi',1)}$ follows from unitarity 
of the $P$-matrix \Erf0P, while the charge conjugation is
  \be  \CO_{(\lambdad,\psi,1),(\lambdad',\psi',1)}
  = \Frac12\, (\Cz_{\lambdad,\lambdad'} + \psi\psi'(P^2)_{\lambdad,\lambdad'})
  \labl{cons}
with
  \be  \Cz_{\lambdad,\lambdad'}
  := \sumeq\mu \Sz_{\lambdad,\mu} \Se_{\mu,\lambdad'} \,.  \ee
We conclude that $\Cz$ must be a permutation of order two; it is convenient 
to use $\Cz$ to define a conjugation on dotted indices, which we denote 
by a superscript `$+$'. Moreover, $P^2$ must be the same
permutation, up to a sign that can depend on $\lambdad$:
  \be  (P^2)_{\lambdad,\lambdad'}^{}
  = \epsd_\lambdad^{}\, \Cz_{\lambdad,\lambdad'} \,.  \ee
{}From \erf{cons} we then learn that
  \be  \CO_{(\lambdad,\psi,1),(\lambdad',\psi',1)}
  = \Cz_{\lambdad,\lambdad'} \,\delta_{\psi,\psi'\epsd_\lambdad}^{} \,.
  \Labl,4
We can determine the signs $\epsd_\lambdad$ from the requirement that 
a consistent conjugation relates fields with identical conformal weight. 
The two conformal weights that have to coincide are $(\Tz_\lambda)^{1/2}\psi$
and $(\Tz_\lambdadp)^{1/2}\psi'\eq(\Tz_\lambdadp)^{1/2}\psi\epsd_{\lambdad}$,
so that
  \be  \epsd_\lambdad = \fraC{(\Tz_{\lambdad})^{1/2}_{}}
  {(\Tz_\lambdadp)^{1/2}_{}} \,,  \Labl,2
which is indeed a sign, and which allows us to write
  \be  P^2 = (\Tz)^{1/2}_{} \Cz (\Tz)^{-1/2}_{} \,.  \Labl pc
Thus the presence of the signs $\epsd_\lambdad$ is
due to the fact that we choose the two square roots
for $\lambdad$ and $\lambdadp$ independently. 
As an easy consequence of the relation \Erf,2 we have
  \be \epsd_\lambdadp^{} = \epsd_\lambdad^{} \,,  \ee
which in turn implies that the conjugation has order two. 

We can summarize our results about charge conjugation as
  \be  \bearl
  (\mu,0,0)^+_{\phantom I} = (\mu^+,0,0)\,,
  \\{}\\[-.8em]
  (\mu,\psi,0)^+_{\phantom I} = (\mu^+,\eps_\mu\psi,0)\,,
  \\{}\\[-.8em]
  (\mud,\psi,1)^+_{\phantom I} = (\mud^+,\epsd_\mud\psi,1)\,.
  \end{array}\ee

As a side remark, let us also mention that in every conformal field theory, 
the requirements that $S^2\eq C$ and $C^2\eq\one$ imply that $S^{-1}\eq SC$,
while the requirements that $S$ is unitary and symmetric imply that
$S^{-1}\eq S^*$; thus charge conjugation and complex conjugation 
of S-matrix elements are related as $S_{i^+,j}^{}\eq(S_{i,j})^*_{}.$
Let us check that this property is indeed satisfied for the orbifold. When 
non-symmetric fields
are involved, the property is obeyed trivially. {}From the diagonal elements in
the twisted sector we find the condition
  \be  P_{\mud,\mud'}^* = \epsd_\mud\, P_{\mud^+,\mud'} \,,  \labl{103}
while the off-diagonal elements yield
  \be (\Se_{\lambda,\mud})^* = \eta_\lambda^{-2}\, \Se_{\lambda,\mud^+}
  = \Sz_{\mud^+,\lambda} \,,  \Labl,3
and
  \be  (\Se_{\lambda,\mud})^* = \eps_\lambda^{} \eta_{\lambda}^*
  \eta_{\lambda^+}^{-1}\, \Se_{\lambdap,\mud} \,.  \Labl48
By unitarity of $\Se$ and by the definition of the conjugation in the
twisted sector, these relations are satisfied automatically.

\subsubsection{The relation between $\SO$ and $\TO$}

Finally we demand that the modular group relation $(\SO\TO)^3\eq(\SO)^2$ 
holds or, equivalently, that
$\TO{}^*\SO\TO{}^*\eq\SO\TO\SO\,{=:}\,\SOTOSO$. This is indeed
the case; for $\SOTOSO_{(\lambda,0,0),(\lambdad',\psi',1)}$ it is
immediate (these matrix elements are zero), and for
$\SOTOSO_{(\lambda,0,0),(\lambda',0,0)}$ it follows directly from 
$(ST)^3\eq S^2$. For $\SOTOSO_{(\lambdad,\psi,1),(\lambdad',\psi',1)}$,
validity of the condition is checked by also using that
$(\Te)^2\eq T$, while for $\SOTOSO_{(\lambda,0,0),(\lambda',\psi,0)}$
and $\SOTOSO_{(\lambda,0,0),(\lambda',\psi,0)}$,
one has to employ the identity
  \be  \sum_{\psi=\pm1} \sumeqd\mu\Se_{\lambda,\mud}\,\TO_{(\mud,\psi,1)}\,
  \Sz_{\mud,\lambda'} = 0 \,,  \Labl0p
which is a consequence of $\TO_{(\lambdad,-\psi,1)}
\eq{-}\TO_{(\lambdad,\psi,1)}$.
Finally, for `mixed' matrix elements involving both untwisted symmetric and
twisted fields, one has to compare
  \be  (\TO{}^*\SO\TO{}^*)^{}_{(\lambda,\psi,0),(\lambdad',\psi',1)}
  = \Frac12\, \psi\psi'\, \eta_\lambda^{-1}\, T_\lambda^*
  \Se_{\lambda,\lambdad'} (\Tz_{\lambdad'})^{-1/2}_{}  \ee
to
  \be  (\SO\TO\SO)^{}_{(\lambda,\psi,0),(\lambdad',\psi',1)}
  = \Frac12\,\psi\psi'\,\eta_\lambda^{-1}\dsty\sumeqd\mu \Se_{\lambda,\mud}
  (\Tz_\mud)^{1/2} P_{\mud,\lambdad'}^{} \,.  \ee
Equality holds if and only if
  \be  \Se (\Tz)^{1/2}_{} P = T^{-1} \Se (\Tz)^{-1/2}_{} \,;  \ee
using the definition of $P$, this is in turn equivalent to the identity
\Erf pc.
We conclude that the requirement $(\SO\TO)^3\eq(\SO)^2$ does not lead to
any new constraints on $\SO$ or $\TO$.

\subsubsection{Collection of the results}

We summarize the findings above by the statement that the consistency
conditions \Erf ST for the matrices $\SO$ and $\TO$ are equivalent to 
the following set of properties of the matrices $\Se$, $\Sz$ and $\Tz$:
  \be\begin{array}{ll}
  (i)   & \Se \;\ \mbox{is unitary}\,, \\[.6em]
  (ii)  & \Sz_{\lambdad,\lambda'} = \eta_{\lambda'}^{-2}\,
          \Se_{\lambda',\lambdad}\,, \\{}\\[-.46em]
  (iii) & \Ce_{\lambda,\lambda'} := \eta_\lambda^{-1} \eta_{\lambda'}^{}
          \, (\Se\Sz)_{\lambda,\lambda'}
        = \eps_\lambda^{} C_{\lambda,\lambda'}^{}\,, \\{}\\[-.6em]
  (iv)  & \Cz_{\lambdad,\lambdad'}:= (\Sz\Se)_{\lambdad,\lambdad'}
        = \Frac{(\Tz_\lambdad)^{1/2}_{}}{(\Tz_{\lambdad'})^{1/2}_{}}\,
          (P^2)_{\lambdad,\lambdad'}^{}\,.
  \end{array}\Labl iv
Here $P$ is defined by \Erf0P; also, the numbers $\eps_\lambda$ 
must all be $\pm1$, and $\Cz$ 
must be a permutation of order two. The requirement $(iv)$ actually 
consists of two conditions, namely that $\Sz\Se$
is a permutation of order two, and that up to signs (which are, however,
uniquely determined) $P^2$ is that same permutation; also, $(iv)$
is the only condition that constrains
the conformal weights in the twisted sector. 

We now draw further conclusions from the constraints \Erf iv.
First, using unitarity of $\Se$ and $\Sz$, we get from condition
$(iv)$ that
  \be  \Se_{\lambda,\lambdad'{}^+_{\phantom|}}
  = (\Sz_{\lambdad',\lambda})^*_{}
  \equiv \eta_\lambda^2\, (\Se_{\lambda,\lambdad'})^*_{} \,;  \ee
this can be regarded as an analogue of simple current symmetries 
\cite{scya} of S-matrices. Second, combining conditions $(ii)$ 
and $(iii)$ with the unitarity of $\Sz$, we learn that
  \be  \Sz_{\mud,\lambdap} = \eps_\lambda^{} \eta_\lambdap^{-1}
  \eta_\lambda^{-1} \cdot (\Sz_{\mud,\lambda})^*_{} \,. \labl{ccond'}
In particular, the relation
  \be  \Sz_{\mud,\lambdap} = (\Sz_{\mud,\lambda})^*_{}  \labl{ccond}
is equivalent to having
  \be  \eps_\lambda^{} = \eta_\lambdap \eta_\lambda^{}\,.  \ee
In this latter case it follows e.g.\ that for every selfconjugate 
$\lambda$ the number $\eta_\lambda^2$ is a sign.
(In particular, when $\gs$ is itself charge conjugation, so that we are
dealing here only with selfconjugate $\lambda$, then $\eta_\lambda^2\eq
{\pm}1$ in full generality.)

\subsection{Fusion rules}

The fusion rules of the orbifold are expressible through $\SO$ via the
Verlinde formula. By direct computation we arrive at the following results.
First, there is the expected twist selection rule
  \be
  \No{(\lambdad_1,\psj_1,g_1)}{(\lambdad_2,\psj_2,g_2)}{(\lambdad_3,\psj_3,g_3)}
  = 0 \quad {\rm for}\;\ g_1{+}g_2{+}g_3\eq1 \bmod 2 \,.  \ee
Here we introduced the convention that $\psj$ can take the values $\pm1$
for $g\eq1$, while for $g\eq0$ the allowed values are $\pm1$ for symmetric
fields, but $0$ for non-symmetric ones.

Also, in the untwisted sector we find linear combinations
of the original fusion rule coefficients as long as
non-symmetric fields are involved:
  \be\begin{array}{ll}
  \No{(\lambda_1,0,0)}{(\lambda_2,0,0)}{(\lambda_3,0,0)}
  &= \N{\lambda_1}{\lambda_2}{\ \lambda_3}
  + \N{\gs\lambda_1}{\lambda_2}{\;\ \ \ \lambda_3}
  + \N{\lambda_1}{\gs\lambda_2}{\;\ \ \ \lambda_3}
  + \N{\lambda_1}{\lambda_2}{\gs\lambda_3} \,,
  \\{}\\[-.7em]
  \No{(\lambda_1,0,0)}{(\lambda_2,0,0)}{(\lambda_3,\psi,0)}
  &= \N{\lambda_1}{\lambda_2}{\ \lambda_3}
  + \N{\lambda_1}{\gs\lambda_2}{\;\ \ \ \lambda_3} \,,
  \\{}\\[-.7em]
  \No{ (\lambda_1,0,0)}{(\lambda_2,\psi_2,0)}{(\lambda_3,\psi_3,0)}\!\!\!
  &= \N{\lambda_1}{\lambda_2}{\ \lambda_3} \,.
  \end{array}\ee
In particular, in these cases the fusion rules of the orbifold theory 
are manifestly non-negative integers. Note that
only fusion rule coefficients of the original
theory appear; this is due to the fact that when at least
one non-symmetric field involved, then the twisted sector does not contribute
to the Verlinde sum. A more interesting case is the remaining one in the 
untwisted sector, which is
  \be
  \Nol{(\lambda_1,\psi_1,0)}{(\lambda_2,\psi_2,0)}{(\lambda_3,\psi_3,0)}
  = \Frac12\, \llb \Nl{\lambda_1}{\lambda_2}{\lambda_3}
  + \psi_1 \psi_2 \psi_3\, ({\eta_{\lambda_1}\eta_{\lambda_2}\eta_{\lambda_3}}
  %% /{\eta_\vac}
  )^{-1}_{}\, \Nel{\lambda_1}{\lambda_2}{\lambda_3} \lrb  \ee
with
  \be  \Nel{\lambda_1}{\lambda_2}{\lambda_3}
  := \sum_\mud \frac{\Se_{\lambda_1,\mud} \Se_{\lambda_2,\mud}
  \Se_{\lambda_3,\mud}} {\Se_{\vac,\mud}} \,.  \Labl51
These coefficients tell us how the chiral blocks of the original theory 
split under the orbifold action. In the case of $\zet_2$-permutation orbifolds
\cite{bohs}, this just amounts to symmetrization or antisymmetrization of
the blocks, so that there is an immediate formula for the numbers
${\rm N}\Untw$. 
In general, we encounter the following structure. An action of the orbifold 
group can not only be defined on the \chir-modules, but also on the spaces 
of chiral blocks of the \chir-theory. (Using the description of chiral blocks
in terms of co-invariants, see e.g.\ \cite{tsuy,beau}, this can be made 
explicit in the case of \wzwts.) The expression appearing
in formula \erf{51} is then the trace
of this action or, to be more precise, since this action is only defined
up to a sign, the difference of the contributions from the
two invariant subspaces. Notice that
the form of \erf{51} is precisely the one that is familiar from the Verlinde
formula. We would also like to point out that simple current symmetries
can be implemented in a similar way on the spaces of chiral blocks
\cite{fuSc8}; in that case a formula for the traces has been conjectured
that is of Verlinde form as well, but with the matrix $\Se$ in the 
denominator of \Erf51 replaced by $S$ \cite{fuSc8}.

For the fusion rules that involve two twist fields we find
  \be
  \Nol{(\lambdad_1,\psi_1,1)}{(\lambdad_2,\psi_2,1)}{(\lambda_3,0,0)}
  = \sumeq\mu \eta_\mu^2\,
  \Frac{\Sz_{\lambdad_1,\mu} \Sz_{\lambdad_2,\mu}
  S_{\lambda_3,\mu}^{\phantom|}} {S_{\vac,\mu}} \labl{101}
(independently of the values of $\psi_1$ and $\psi_2$) and
  \be \Nol{(\lambdad_1,\psi_1,1)}{(\lambdad_2,\psi_2,1)}{(\lambda_3,\psi_3,0)}
  = \Frac12\, \llb \sum_\mu \eta_\mu^2\,
  \Frac{\Sz_{\lambdad_1,\mu} \Sz_{\lambdad_2,\mu} S_{\lambda_3,\mu}^{}}
  {S_{\vac,\mu}} + \psi_1\psi_2\psi_3 \sum_\mud \eta_{\lambda_3}^{-1}\,
  %%\eta_\vac^{}\,
  \Frac{P_{\lambdad_1,\mud}^{} P_{\lambdad_2,\mud}^{} \Se_{\lambda_3,\mud}}
  {\Se_{\vac,\mud}} \lrb \,.  \labl{102}
Note that upon summation over $\psi_3$ the second term in \erf{102} cancels
out, so that we have 
  \be  \sum_{\psi_3=\pm1} 
  \Nol{(\lambdad_1,\psi_1,1)}{(\lambdad_2,\psi_2,1)}{(\lambda_3,\psi_3,0)}
  = \sumeq\mu \eta_\mu^2\,
  {\Sz_{\lambdad_1,\mu} \Sz_{\lambdad_2,\mu} S_{\lambda_3,\mu}}/\,
  {S_{\vac,\mu}} \,,  \ee
which is of the same form as the expression \erf{101}
for $\Nol{(\lambdad_1,\psi_1,1)}{(\lambdad_2,\psi_2,1)}{(\lambda_3,0,0)}$.
In the last few equations we have presented the structure constants with three
lower indices. Those with an upper index are obtained by either using a
complex conjugate matrix $(\SO)^*_{}$ in the Verlinde formula or,
equivalently,\,%
 \futnote{When deriving \Erf,0 this way from \erf{101}, in addition to the 
conjugation \Erf,4 one has to employ the identity \erf{ccond'}.}
by raising the index with the help of the conjugation matrix $\CO$ of the
orbifold; e.g.\ one has
  \be  \No{(\lambdad_2,\psi_2,1)}{(\lambda_3,0,0)}{(\lambdad_1,\psi_1,1)} 
  = \sumeq\mu 
  \Frac{(\Sz_{\lambdad_1,\mu})^*_{\phantom|} \Sz_{\lambdad_2,\mu}
  S_{\lambda_3,\mu}^{\phantom|}} {S_{\vac,\mu}}\,.  \labl,0

It is readily checked that the orbifold field
  \be  \J\orb := (0,-1,0)  \ee
acts under fusion as
  \be  \J\orb \star (\lambda,\eps,g) = (\lambda,-\eps,g)  \ee
and hence is a simple current of order two. It has integral conformal weight, 
and the extension of the orbifold theory by this simple current simply
reproduces the original theory. Upon extension, orbifold
fields coming from symmetric fields of the original theory
form full orbits, whereas the fields coming from non-symmetric fields
are fixed points which need to be resolved, thus giving rise to a
pair of non-symmetric fields.
The virtue of this simple relationship is that it provides us with two
distinct descriptions of one and the same situation -- a pair of theories
with respective chiral \alg s \chir\ and $\chir\orb{\subset}\,\chir$.
Comparison of the two descriptions often simplifies the analysis of this 
situation. This was instrumental in the investigation of \bc s in
\cite{fuSc10,fuSc11}, e.g.\ when discussing the relationship between
boundary blocks (i.e., chiral blocks for one-point functions of bulk
fields on the disk) and boundary states.

\sect{WZW orbifolds}

In this section we apply the general results of section 2 to the case where
the original theory is a \wzwt\ based on some untwisted affine \lie\ \g. In this
situation every \auto\ of the full chiral \alg\ \chir, which for our 
present purposes is the semi-direct
sum of \g\ and the Virasoro \alg, is completely determined by
its restriction $\omega$ to \g. Moreover,
of particular interest in applications are those cases where this \auto\
$\omega$ comes from an \auto\ of the horizontal sub\alg\ \gb\ of 
$\g\,{\equiv}\,\gb\untw$. 
In the case of interest to us here, i.e.\ orbifolds, this restriction
is mandatory because
the \auto\ must act as the identity on the Virasoro \alg\ and hence must
preserve \gb; accordingly here we consider this particular kind of \auto s
of \g. Also, in order to make contact with the discussion in section 2 we
require $\omega$ to have order two, i.e.\ generate a $\zet_2$-group. For 
the case of {\em inner\/} \auto s, WZW orbifolds \wrt more general finite
orbifold groups have been discussed in \cite{kaTo2}.
In the present paper we use the techniques developped in \cite{fusS3,furs}
which allow us to deal also with the case of outer automorphisms. There are
two reasons to expect that in the case of an outer automorphism of 
$\g\eq\gb\untw$ the twisted
sector can be understood in terms of the {\em twisted\/} affine \lie\
$\gb\tw$. To compute the traces in the untwisted sector, we will use
the theory of twining characters and orbit \lie s as introduced in
\cite{fusS3}, and in the case at hand the latter are 
twisted affine \lie s. Modular transformation of these
traces yields again characters of twisted affine \lie s, although not
necessarily of the same twisted affine \lie\ that is relevant to the
untwisted sector. The fact that the modules
of twisted affine \lie s furnish the states of the twisted sector can,
of course, also be understood from the fact that the latter provide
twisted \rep s of the chiral algebra $\chir$ in the sense of \cite{dolm3}.

\subsection{Automorphisms}

For simplicity, we denote the relevant \auto\ of the \findim\
simple \lie\ \gb\ again by $\omega$. It is known \cite[Prop.\,8.1]{KAc3} 
that for every \auto\ $\omega$ of \gb\ of finite order there is a suitable 
Cartan\hy Weyl basis of \gb\ in which it can be written as
  \be  \omega = \omego \circ \sigmas \,,  \Labl om
where $\omego$ is a diagram automorphism, i.e.\ acts on the Chevalley
generators of \gb\ as
  \be  \omego(E^i)=E^{\dot\omega i} \,, \quad \omego(F^i)=F^{\dot\omega i} \,,
  \quad \omego(H^i) = H^{\dot\omega i}  \Labl32
($i\eq1,2,...\,,r\,{\equiv}\,\rank\g$)
with some (possibly trivial) symmetry $\dot\omega$ of the Dynkin diagram 
of \gb, while $\sigmas$ is an inner automorphism
  \be  \sigmas = \exp(2\pi\ii\, {\rm ad}_{H_s}^{}) \,,  \ee
where $H_s\,{\equiv}\,(s,H)$ is some element of the \csa\ of \gb\ that
satisfies $\omega(H_s)\eq H_s$.
The latter property implies that the \auto s $\omego$ and $\sigmas$
commute. It is not difficult to see that
every automorphism of the form \Erf om extends uniquely to an \auto\ 
of the untwisted affine \lie\ $\g\eq\gb\untw$, which acts like
  \be  \bearll
  \omega(E^i_n)= (\Zeta)^n_{}\,\eE^{2\pi\ii(s,\alpha^{(i)})}
  E^{\dot\omega i}_n \,,\ &
  \omega(F^i_n)= (\Zeta)^n_{}\,\eE^{-2\pi\ii(s,\alpha^{(i)})}
  F^{\dot\omega i}_n \,, \\{}\\[-.5em]
  \omega(H^i_n) = H^{\dot\omega i}_n \,, & \omega(K) = K 
  \,.  \eear \Labl aa
Here $\alpha^{(i)}$ are the simple \gb-roots,
$K$ denotes the canonical central element of \g, and $\Zeta$ is a sign
defined by
  \be  \omego(E^\theta) = \Zeta\,E^\theta \,,  \ee
where $\theta$ is the highest \gb-root. In particular, for $s\eq0$ the generators
coming from the additional simple root $\alpha^{(0)}$ of \g\ transform as
  \be  \omego(E^0)=\Zeta\,E^0 \,, \quad \omego(F^0)=\Zeta\,F^0 \,,
  \quad \omego(H^0) = H^0 \,;  \Labl34
for $\Zeta\eq{+}1$ this is nothing but the diagram \auto\ of \g\
that is obtained by the same prescription as in \Erf32 when one 
extends $\dot\omega$ in the natural manner, i.e.\ as $\dot\omega0\eq0$.
By direct calculation, one finds that\,%
 \futnote{The \auto\ obtained by setting $\Zeta$ to 1 in the exceptional
$A_{2n}$-case does not leave the Virasoro algebra invariant; 
it maps $L_n$ to $(-1)^n_{}L_n$.}
  \be  \Zeta = \left\{\begin{array}{rl} -1 & {\rm for}\
  \gb\eq A_{2n},\; \omego\eq\oc \,, \\[.2em]  1 & {\rm else}\,.  \eear
  \right. \labl{Zeta}

We identify the
\csa\ with the weight space and correspondingly call $s$ the {\em shift
vector\/} that characterizes the inner \auto\ $\sigmas$.
The map $\gs$ on the weight space only depends on $\omego$; its
action on the fundamental \gb-weights $\Lambda_{(i)}$ reads
  \be  \gs \Lambda_{(i)} = \Lambda_{(\dot\omega i)} \,.  \ee
In particular, for inner \auto s $\omega$ the map $\gs$ is the identity.
We refer to \g-weights $\lambda$ that satisfy
$\gs\lambda\eq\lambda$ as {\em symmetric\/} weights; e.g.\ according
to the statements above the shift vector $s$ is symmetric,
  \be  \gs s = s \,.  \Labl,5
Now the group of automorphisms of a \findim\ simple \lie\
is a real compact Lie group; its factor group modulo inner automorphisms
is isomorphic to the center $Z(G)$ of the universal covering group $G$ whose
\lie\ is the compact real form of \gb. 
Moreover, every element of the center $Z(G)$ can be obtained by
exponentiation of an element of the coweight lattice $Q^*$ of \gb.
As a consequence, the shift vector $s$ is only defined
up to a symmetric element of the coweight lattice. The inequivalent 
shift vectors are characterized in theorem 8.6 of \cite{KAc3}.

For $\omega$ to have order two, $2s$ must be an element of the 
coweight lattice $Q^*$, and hence corresponds to a group
element $\gamma_{2s}\eq\eE^{2\pi\ii\cdot2s}$ in the center $Z(G)$. 
According to the results of \cite{scya,scya6} it therefore
determines uniquely a simple current of the \wzwt\ based on \g,
which we denote by $\J^\ssss$. (But the shift vector $s$ is {\em not\/}, 
in general, equal to 1/2 the co-minimal
fundamental weight that characterizes \cite{fuge} the simple current
$\J^\ssss$.) We also introduce the monodromy charge
  \be  Q^{}_{\J^\ssss}(\lambda) := \Delta(\lambda) + \Delta(\J^\ssss)
  - \Delta(\J^\ssss{\star}\lambda) \, \bmod \zet  \ee
of $\lambda$ \wrt the simple current $\J^\ssss$. When multiplied
with the order of $\J^\ssss$, $Q^{}_{\J^\ssss}(\lambda)$ equals
minus the conjugacy class of $\lambda$ with respect to that element 
$\gamma_{2s}\iN Z(G)$ that is obtained as the exponential of the element 
$2s$ of the coweight lattice. 
Thus the exponentiated monodromy charge is given by
  \be  \exp(2\pi\ii Q^{}_{\J^\ssss}(\lambda)) = (\etas_\lambda)_{}^{-2}
  \Labl qs
with
  \be  \etas_\lambda:= \exp(2\pi\ii(s,\lambda)) \,.  \labl{etas}
It follows \cite{scya} in particular that
  \be  S_{\lambda,\mu}^{}\,\etas_\mu{}_{}^2 = S_{\J^\ssss\star\lambda,\mu}  
  \Labl82
for all $\lambda,\,\mu$.

The (conjugacy classes of) order-two \auto s $\omega$ of a complex 
\findim\ simple \lie s are uniquely characterized by their fixed point
\alg\ $\gb^\omega$, which is the subalgebra of \gb\ that is left 
pointwise fixed under $\omega$. Equivalently, these \auto s are in 
one-to-one correspondence to the real forms of these complex \lie s. 
A complete list of the latter can e.g.\ be found in table 
II on p.\ 514 of \cite{HElg}.
 \futnot{compare also table 9.3 of [GIlm].}
It is straightforward (though lengthy) to determine the
corresponding diagram \auto s and shift vectors. We summarize the
pertinent results in table \ref{ta}.
In table \ref{ta} we use the following notation. Except for $\gb\eq D_4$,
$\oc$ is the unique non-trivial diagram \auto\ of the relevant \lie, while
for $\gb\eq D_4$ it is the diagram \auto\ for which $\dot\omega$ exchanges
the two spinor nodes of the Dynkin diagram.
In all cases where the group of simple currents of the theory is cyclic,
$\J$ stands for the generator of this group that has highest weight
$k\Lambda_{(1)}$, except for $\gb\eq C_n$, where the weight is
$k\Lambda_{(n)}$; for $D_n$ we use the notation $\Jv$ for the vector current,
of highest weight $k\Lambda_{(1)}$,
which has order two and whose monodromy charge distinguishes spinor and tensor 
representations, while $\Js$ is the spinor current, of highest weight
$k\Lambda_{(n)}$.
  \begin{table}[bhtp]\caption{{\em Order-2 \auto s of \findim\ simple
  \lie s.}}
  \begin{center}
  $   \begin{array}{|l|c|c|c|c|c|}
  \hline &&&&&\\[-.82em]
  \multicolumn1{|c|} \gb & \omego & s & (s,s) & \J^\ssss & \gb^\omega
  \\[-.97em]{}&&&&&\\\hline\hline&&&&&\\[-.77em]
  A_1     & \id & \frac12\,\Lambda_{(1)} &   1/8   &  \J    & \uone        
          \nxL
  A_{n,\;\ n>1}\! & \id
          & \frac12\,\Lambda_{(\ell)}, \;\ell\eq2,...\,,[\frac{n+1}2]
          & \ell(n{+}1{-}\ell)/4(n{+}1) & \J^{-\ell}
                                            & A_{\ell-1}\oplU A_{n-\ell}\oplU\uone
          \nxL
  A_{2n}  & \oc & 0                      &    0    & \bfe   & B_n
          \nxL
  A_{2n+1}& \oc & 0                      &    0    & \bfe   & C_{n+1}
          \nxl
          & \oc &\frac12\,\Lambda_{(n+1)}&(n{+}1)/8&\J^{n+1}& D_{n+1}
          \nxL
  B_n     & \id & \frac12\,\Lambda_{(\ell)}
                ,\; \ell\eq2,...\,,n{-}1 & \ell/4  &\J^\ell & D_\ell\oplu B_{n-\ell}
          \nxL
 %B_{2n}  & \id & \frac12\,\Lambda_{(n)} &   n/4   &  \J^n  & B_n\oplu D_n
 %        \nxL
 %B_{2n+1}& \id &\frac12\,\Lambda_{(n+1)}&(n{+}1)/4&\!\J^{n+1}\!& D_{n+1}\oplu B_n
 %        \nxL
  C_n     & \id & \Lambda_{(\ell)},\;
                    \ell\eq1,...\,,[n/2] & \ell/2  & \bfe   & C_\ell\oplu C_{n-\ell}
          \nxl
          & \id & \frac12\,\Lambda_{(n)} &   n/8   &  \J    & A_{n-1}\oplu\uone
          \nxL
  D_n     & \id & \frac12\,\Lambda_{(\ell)},
          \;\ell\eq2,...\,,[\frac{n+1}2] & \ell/4  &\Jv^\ell& D_\ell\oplu D_{n-\ell}
          \nxl
          & \id & \frac12\,\Lambda_{(n)} &   n/16  &  \Js   & A_{n-1}\oplu\uone
          \nxl
          & \oc & 0                      &    0    & \bfe   & B_{n-1}
          \nxl
          & \oc & \frac12\,\Lambda_{(\ell)},
          \;\ell\eq1,...\,,[\frac{n-1}2] & \ell/4  &\Jv^\ell& B_\ell\oplu B_{n-\ell-1}
          \nxL
 %D_{2n}  & \id & \frac12\,\Lambda_{(n)} &   n/4   & \Jv^n  & D_n\oplu D_n
 %        \nxL
 %D_{2n+1}& \oc & \frac12\,\Lambda_{(n)} &   n/4   & \Jv^n  & B_n\oplu B_n
 %        \nxL
  E_6     & \id & \frac12\,\Lambda_{(1)} &   1/3   &  \J    & D_5\oplu\uone
          \nxl
          & \id & \frac12\,\Lambda_{(6)} &   1/2   & \bfe   & A_5\oplu\uone
          \nxl
          & \oc & 0                      &    0    & \bfe   & F_4
          \nxl
          & \oc & \frac12\,\Lambda_{(6)} &   1/2   & \bfe   & C_4
          \nxL
  E_7     & \id & \frac12\,\Lambda_{(5)} &    1    & \bfe   & D_6\oplu\uone
          \nxl
          & \id & \frac12\,\Lambda_{(6)} &   3/8   &  \J    & E_6\oplu\uone
          \nxl
          & \id & \frac12\,\Lambda_{(7)} &   7/8   &  \J    & A_7
          \nxL
  E_8     & \id & \frac12\,\Lambda_{(1)} &   1/2   &  \bfe  & E_7\oplu\uone
          \nxl
          & \id & \frac12\,\Lambda_{(7)} &    1    &  \bfe  & D_8
          \nxL
  F_4     & \id & \frac12\,\Lambda_{(1)} &   1/2   &  \bfe  & C_3\oplu A_1
          \nxl
          & \id &          \Lambda_{(4)} &    1    &  \bfe  & B_4
          \nxL
  G_2     & \id & \frac12\,\Lambda_{(1)} &   1/2   &  \bfe  & A_1\oplu A_1
  \\[.47em]\hline 
  \end{array}$ \label{ta} \end{center} \end{table}

By comparing the results for the numbers $(s,s)$ with the values \cite{fuge} of
the conformal dimensions $\Delta(\J)$ of simple currents, one finds that
the identity
  \be  \Delta(\J^\ssss)/k = 2\,(s,s) \,\bmod \zet  \Labl02
holds for all entries in table \ref{ta}. Together with the result \Erf qs and 
the identity
  \be  Q_\J(\J{\star}\lambda) +Q_\J(\lambda) = 2\,\Delta(\J) \,\bmod \zet
  \,,  \ee
this implies that
  \be  (\etas_\lambda\, \etas_{\J^\ssss\star\lambda})_{}^2
  = \eE^{-4\pi\ii\Delta(\J^\ssss)} = \eE^{-8\pi\ii k(s,s)}  \Labl81
   for every $\lambda$.

\subsection{Implementation of automorphisms}

To be able to determine the characters of the orbifold, 
we have to know the action of the automorphism $\omega$ on the primary
fields. To this end we implement $\omega$ by suitable maps
$\Tau_\lambda$ on \ihwm s $\calh_\lambda$ of the affine \lie\ \g.\,%
 \futnote{By a slight abuse of notation, we use the symbol $\lambda$ both
for a highest weight of \gb\ and for a highest weight of \g. This is justified
by the fact that (as in any application to \cft) we work at a fixed
non-negative integral value of the level of the \g-modules, 
so that the \gb-weight uniquely determines its affine extension.}
They must satisfy
the twisted intertwining property $\Tau_\lambda\,{\circ}\,x\eq\omega(x)\,
{\circ}\,\Tau_\lambda$ for all $x\iN\g$. This requirement determines them 
up to a scalar factor. One possible implementation is
  \be  \tilde\Tau_\lambda := \exp(2\pi\ii (s,H))\, \Tauo_\lambda  \,.  \Labl i1
Here $\Tauo_\lambda$ is the preferred implementation of the diagram
automorphism, i.e.\ the one that acts as the identity map on the highest
weight space of $\calh_\lambda$.\,%
 \futnote{This implementation has been used in \cite{fusS3,furs}.}
The inner part, $\exp(2\pi\ii(s,H))$,
can be regarded as the representation matrix $R_\lambda(\gamma^\ssss)$
of an element $\gamma^\ssss$ of the simply connected,
connected, compact Lie group whose Lie \alg\ is the compact real form of \gb.
Unlike $\omega$ itself, the twisted intertwiners \Erf i1 do not square to the
identity map. Rather $(\Tau_\lambda)^2_{}$ acts on
$\calh_\lambda$ as a multiple $(\etas_\lambda)^2_{}\,\id$ of the identity, where
$\etas_\lambda$ is the number introduced in \erf{etas}.
An implementation which does have order two is thus given by
  \be  \Tau_\lambda := (\etas_\lambda)^{-1}_{}\,\tilde\Tau_\lambda \,.  \Labl1i
It should be realized that the requirement that the map
$\Tau_\lambda$ has order two
does not fix it uniquely, but leaves an over-all sign undetermined; in the
prescription \Erf1i we have fixed this ambiguity by requiring that the map
acts as $+\id$ on the highest weight vector.
 \futnot{While this looks natural
in the WZW case, it should be kept in mind that this is {\em not\/}
the only way to fix $\Tau_\lambda$, and in fact it is not even natural.}

Recall that $\etas_\lambda{}_{}^2$ is the monodromy charge 
\wrtt simple current $\J^\ssss$; thus the number $\etas_\lambdap{}_{}^2
\eq\exp(2\pi\ii(2s,\lambdap))\eq\exp(2\pi\ii(2s^+,\lambda))$ 
is the monodromy charge \wrtt inverse of that simple 
current, so the product of these two numbers is 1. It follows that
  \be  \epss_\lambda := \etas_\lambda\etas_\lambdap \Labl pm
is equal to $\pm 1$.

Our goal is now to compute the characters of the primary fields in the
untwisted sector of the orbifold. While the identification of pairs
of non-symmetric fields does not pose any problem, in the symmetric case we 
need to get a handle on the
eigenspaces of the action of $\Tau_\lambda$ on the modules having symmetric 
highest weight. Using the projectors 
  \be  \bfe+\psi\, \Tau_\lambda \ee
on the eigenspaces, where $\psi$ takes the values $\pm1$, 
the characters of the orbifold primaries in the untwisted sector are
  \be  \chii_{(\lambda,\psi,0)}(\tau,z)
  = \Frac12\, \Tr_{\calh_\lambda} (\bfe\,{+}\,\psi\, \Tau_\lambda)\,
  q^{L_0-c/24} \eE^{2\pi\ii (z,H)} 
  = \Frac12\, \chii_\lambda(\tau,z) + \Frac12\, {(\etas_\lambda)}^{-1} \psi\,
  \chio_\lambda(\tau,z) \,. \labl{c1}
Here we introduced the expression
  \be \chio_\lambda(\tau,z):= \Tr_{\calh_\lambda}\tilde\Tau_\lambda\,
  q^{L_0-c/24} \eE^{2\pi\ii (z,H)} \,.  \ee
for every automorphism $\omega$ of \g; these
quantities $\chio_\lambda$ are known \cite{fusS3}
as {\em twining characters\/} for the automorphism $\omega$. They can be
regarded as a character valued indices.

In the case of diagram automorphisms, the twining characters have been
computed in \cite{fusS3,furs}. It has been proven that, upon correctly
adjusting its arguments, the twining character coincides with the ordinary 
character of some other \lie, which is called the orbit \lie. 
One key ingredient in the proof of this assertion
was the observation that the subgroup of the Weyl group of
\g\ that commutes with the action of the automorphism on the weight space
is isomorphic to the Weyl group of the orbit \lie. The orbit \lie\ is
a Kac\hy Moody algebra of the same type as the original algebra. Thus in 
the case at hand, it is still an affine \lie, albeit no longer necessarily an 
untwisted one; indeed it is a {\em twisted\/} affine \lie\ precisely if the 
\auto\ $\omega$ of \g\ is outer. The
appearance of twisted affine \lie s for those twining characters will 
imply, after modular transformation, that the twisted sector of the orbifold
theory is controlled by the \rep\ theory of a twisted affine \lie. This does
not come as a surprise, since twisted affine \lie s provide twisted \rep s
for untwisted affine \lie s.
For inner automorphisms the orbit \lie\ coincides with \g, so in particular
its horizontal subalgebra is just the horizontal subalgebra \gb\ of \g. For
the outer automorphisms the horizontal subalgebra of the orbit \lie\ is
$C_n$ for $\gb\eq A_{2n}$, $B_{n+1}$ for
$\gb\eq A_{2n+1}$, $C_{n-1}$ for $\gb\eq D_n$ and $F_4$ for $\gb\eq E_6$.

\subsection{Weyl groups and lattices}\label{su.wl}

As we will see below, for orbifolds \wrt\ inner \auto s, the Weyl group $\W$ 
of \g\ plays an analogous role for the twining characters as it does for
the characters of the original \wzwt. In contrast, in the outer case, where
the map $\gs$ on the weight space is non-trivial, this role is taken over by 
the commutant 
  \be  \Wo := \{ \w\iN\W \,|\, \w{\circ}\gs\eq\gs{\circ}\w \}  \ee
of $\gs$ in $\W$. This subgroup $\Wo$ can be described more explicitly as follows.
The Weyl group $\W$ has the structure $\W\eq\Wb\,\Ltimes\,\LV$
of a semi-direct product of the Weyl group $\Wb$ of the horizontal 
subalgebra \gb\ with the coroot lattice $\LV$ of \gb. Thus every element
$\w\iN\W$ can be uniquely written as
  \be  \w = \wb \circ t_\beta \labl{expr}
for some $\beta\iN\LV$, where $t_\beta$ stands for translation by $\beta$; 
conversely, all these maps are in the Weyl group $\W$.
Now since $\omega$ is also an automorphism of \gb, the product 
$\wb_\omega\df(\gs)^{-1} \wb \gs$ is again an element of $\Wb$.
Moreover, $\gs$ restricts to an automorphism of the
coroot lattice $\LV$, and hence
  \be  \w \circ \gs = \wb \circ t_\beta \circ \gs
  = \gs \circ \wb_\omega \circ t_{(\gs)^{-1}_{}\beta}\,. \ee
As the decomposition \erf{expr} is unique, this implies that
$\gs$ commutes with $\w$ if and only if $\wb_\omega\eq\wb$ and
$\gs\beta\eq\beta$. It follows that $\Wo$ is the semi-direct product 
  \be  \Wo = \Wob \;\Ltimes\; \LVo  \ee
of the symmetric part $\LVo$ of the coroot lattice of \gb\ with 
the group
  \be  \Wob := \{ \wb\iN\Wb \,|\, \wb{\circ}\gs\eq\gs{\circ}\wb \} \,.  \ee
It is known \cite{fusS3,furs,muhl} that $\Wob$ is a Coxeter group;
we will denote by $\epso$ its sign function. In fact, $\Wob$ is
nothing but the Weyl group of the orbit
\lie\ \cite{fusS3,furs} of the \findim\ simple \lie\ \gb.

A general element of $\LVo$ is of the form\,%
 \futnote{In all cases with non-trivial $\omego$, \gb\ is simply
laced, so that the root lattice and coroot lattice coincide.}
  \be \sum_{i=1}^r n_i \alpha^{(i)} \quad{\rm with}\;\
  n_i\iN\zet\;\ {\rm and}\;\ n_{\dot\omega i}\eq n_i \quad{\rm for\
  all}\;\ i \,.  \ee
In other words, the basis elements of $\LVo$ are associated to
orbits of $\dot\omega$, and for length-1 orbits the basis element is
just $\alpha^{(i)}$, while for length-2 orbits it is
$\alpha^{(i)}{+}\,\alpha^{(\dot\omega i)}$.
By computing the inner products between these basis vectors,
we find that these lattices are just scaled root lattices,
as listed in the following table.
  \be\begin{array}{|l|c|l|r|}
  \hline &&&\\[-.95em]
  \multicolumn1{|c|} \gb & \gb^\omego & \multicolumn1{c|} \LVo
  & \multicolumn1{c|} \MVo \\[-.97em]{}&&&\\\hline\hline&&&\\[-.81em]
  A_{2n}   &  C_n & \sqrt2\, Q(B_n) &\frac1{\sqrt2}\,Q(B_n)
  \\{}&&&\\[-.83em]
  A_{2n-1} &  B_n & \sqrt2\, Q(B_n)               &  Q(C_n)     
  \\{}&&&\\[-.79em]
  D_{n+1}  &  C_n & \sqrt2\,Q(C_n)\equiv Q(D_n)\! &  Q(B_n)
  \\{}&&&\\[-.79em]
  E_6      &  F_4 & \sqrt2\, Q(F_4)               &  Q(F_4) 
  \\[.3em] \hline \end{array} \Labl tb

Table \Erf tb also provides another set of lattices, denoted by $\MVo$;
they are constructed in the following manner.
The lattice $\LVo$ is the `symmetric' (i.e.\ pointwise fixed) sublattice 
of the coroot lattice $\LV$. We will also need the
symmetric sublattice of $\Lw$, which we denote by $\Mwo$. When $\omego$
is non-trivial, then 
unlike $\LV$ and $\Lw$ themselves, these are no longer dual to each
other. Rather, the dual lattice $\Lwo$ of $\LVo$ is a lattice that contains 
the symmetric weight lattice $\Mwo$ as a sublattice, while 
the dual lattice ${(\Mwo)}^*\,{=:}\,\MVo$ contains the symmetric 
coroot lattice $\LVo$; thus
  \be  \LVo \subseteq \MVo \,, \qquad \Mwo \subseteq \Lwo \,.  
  \Labl in
All these lattice have the same rank, which we denote by $r_\omego$.
Denoting the length of the $\dot\omega$-orbit through $i$,
which is either 1 or 2, by $\ell_i$, we have explicitly
  \be  \bearl
  \LVo = \{ \sum_{i=1}^r n_i\, \alpha^{(i)\Vee} \,|\,
  n_{\dot\omega i}\eq n_i\iN\zet \} \,,
  \\{}\\[-.5em]
  \MVo = \{ \sum_{i=1}^r n_i\, \alpha^{(i)\Vee} \,|\,
  \ell_i n_i\iN\zet,\; n_{\dot\omega i}\eq n_i \} \,,
  \\{}\\[-.5em]
  \Lwo = \{ \sum_{i=1}^r \lambda^i\Lambda_{(i)} \,|\,
  \ell_i\lambda^i\iN\zet,\; \lambda^{\dot\omega i}\eq\lambda^i \} \,,
  \\{}\\[-.5em]
  \Mwo = \{ \sum_{i=1}^r \lambda^i\Lambda_{(i)} \,|\,
  \lambda^{\dot\omega i}\eq\lambda^i\iN\zet \} \,.  \eear \Labl LM
The inclusions \Erf in are both of finite index. In fact, the indices
are equal. Namely, there are the isomorphisms
  \be  \Mwo/\LVo  \cong (\Lwo/\MVo)^*_{} \cong \Lwo/\MVo  \labl{1.5}
of finite abelian groups
(the first isomorphism is canonical, while the second is a non-canonical
isomorphism between a finite abelian group and its dual group), which
implies in particular that
  \be  |\MVo/\LVo| = |\Lwo/\Mwo| \,.  \labl{1.6}

As a sublattice of the coroot lattice $\LV$, which is an even lattice,
$\LVo$ is even as well. Moreover, except for $\gb\eq A_{2n}$ with outer
\auto, the lattice $\sqrt2\MVo\,{\supseteq}\,\sqrt2\LVo$ is also an
even lattice. This can be read off table \Erf tb above, where the 
lattices $\MVo$ are listed, but also follows from the following general
argument. First, every $\betad\iN\MVo$ can be written as
  \be  \betad = \sum_{i=1}^r \Frac{m_i}{\ell_i}\, \alpha^{(i)\Vee}
  = \sum_{i=\dot\omega i} m_i\, \alpha^{(i)\Vee} + \Frac12
  \sum_{i<\dot\omega i} m_i\, (\alpha^{(i)\Vee}{+}\,\alpha^{(\dot\omega i)\Vee})
  \ee
with $m_i\iN\zet$ for all $i\eq1,2,...\,,r$.
Using the symmetry property $A^{\dot\omega i,\dot\omega j}\eq A^{i,j}$
of the Cartan matrix $A$ of \gb\ \wrt $\dot\omega$, it follows that
  \be  (\betad,\betad')
  = \Frac12\!\! \sum_{\scs i,j \atop\scs i<\dot\omega i,\,j<\dot\omega j} 
  \!\!\! m_i^{} m'_j\,
  (\alpha^{(i)\Vee}\!,\alpha^{(j)\Vee}{+}\,\alpha^{(\dot\omega j)\Vee})
  +\!\! \sum_{\scs i,j \atop\scs i=\dot\omega i\;{\rm or}\;j=\dot\omega j}
  \!\!\! m_i^{} m'_j\, (\alpha^{(i)\Vee}\!,\alpha^{(j)\Vee})
  \,\in\Frac12\,\zet\,.  \ee
Moreover, specializing to $\betad'\eq\betad$ and using the explicit form of
the Cartan matrix, one verifies that
  \be  (\betad,\betad) \in \left\{\begin{array}{rl} \Frac12\,\zet & {\rm for}\ 
  \gb\eq A_{2n},\; \omego\eq\oc \,, \\[.2em]  \zet & {\rm else}\,.  \eear
  \right. \Labl37

To relate vectors in the various lattices, we introduce the linear map
  \be  \sigm:\quad \lambda\eq\sum_i\lambda^i\Lambda_{(i)} \;\mapsto\;
  \sigm(\lambda)\df \sum_i \Frac1{\ell_i}\, \lambda^i\Lambda_{(i)} \ee
on the weight space of \gb; this map restricts to bijections
  \be  \sigm(\LVo) = \MVo\,, \qquad \sigm(\Mwo) = \Lwo  \ee
(they are {\em not\/} isomorphisms of lattices; e.g.\ they are not 
isometries).
Since the elements of the Weyl group $\Wb$ of \gb\ preserve the weight
lattice $\Lw$ as well as the coroot lattice $\LV$, it follows that
the elements of $\Wob$ preserve both $\Mwo$ and $\LVo$. Moreover, as elements
of $\Wb$, they are manifestly isometries, and hence they also preserve the
dual lattices $\MVo$ and $\Lwo$. In short, for all the lattices \Erf LM
the elements of $\Wob$ are lattice-preserving isometries. Further, the action 
of $\Wo$, and hence also of $\Wob\,\Ltimes\,h\LVo$ for any $h\iN\zetplus$
-- and similarly also the action of $\Wob\,\Ltimes\,h\MVo$ -- partitions 
the symmetric subspace of the weight space of \gb\ into chambers. 
In order to account for this feature, we introduce the following notation. 
The orbit spaces for the two actions are denoted by 
  \be  \Po_h := \Mwo/(\Wob\Ltimes h\LVo) \qquad\mbox{and}\qquad
  \Pd_h := \Lwo/(\Wob\Ltimes h\MVo) \,,  \labl{Pd} 
\resp. In both cases the subset of orbits on which the action is free plays
a particularly important role. We denote them by 
  \be \PPo_k := (\Po_{k+\gv})^\circ_{} \qquad\mbox{and}\qquad
  \PPd_k := (\Pd_{k+\gv})^\circ_{} \,,  \ee
where for later convenience we have shifted the index by the dual Coxeter
number $\gv$ of \g. 

In the case of undotted weights, natural representatives for $\Po_h$
and $\PPo_k$ exist:
  \be  \bearll
  \Po_h \!\! 
  &= \{ \lambda\,{=}\sum_{i=1}^r \lambda^i\Lambda_{(i)} \,|\, \lambda^
  {\dot\omega i}\eq\lambda^i\iN\zetpluso,\; (\lambda,\theta^\Vee){\le}h \}
  \,, \\{}\\[-.6em]
  \PPo_k \!\!
  &= \{ \lambda\,{=}\sum_{i=1}^r \lambda^i\Lambda_{(i)} \,|\,
  \lambda^{\dot\omega i}\eq\lambda^i\iN\zetplus,\; (\lambda,\theta^\Vee){<}
  {k{+}\gv} \} \,.
  \end{array} \Labl,;
Notice that in general it is {\em not\/} true that $\PPd_h$ coincides with 
$\sigm(\PPo_h)$ (the latter does hold, however, when the map $\sigm$ commutes 
with the action of $\Wob$ which, besides the trivial case where $\omega$ is 
inner, happens if and only if $\gb\eq\A_{2n}$).
But we will see later (see formula \erf{chii'} below) that the set
$\PPd_h$ is closely related to the set of integrable weights of twisted
affine \lie s.

It should also be noted that from their definition it is not obvious
(except when $\sigm$ commutes with $\Wob$)
whether the cardinalities of the two sets $\Po_k$ and $\Pd_k$ are the same.
But it follows from the unitarity of the matrix $\Breve S$ that will be
introduced in \Erf46 below that this equality indeed holds true:
  \be  |\Pd_k| = |\Po_k|  \ee
for all horizontal \alg s \gb, all \auto s $\omega$ and all levels $k$.

\subsection{Twisted characters}

We are now ready to introduce the relevant {\em twisted characters\/} that
appear as constituents of the orbifold characters. They are functions which
generalize the ordinary characters of affine \lie s, and accordingly they
depend on two 
arguments, a number $\tau$ in the upper complex half plane and an element 
$z$ of the weight space $\LV{\otimes_\zet^{}}\reals$ of \gb, as well as
on two parameters, a non-negative integer $h$ and an element of the 
relevant dual lattice, that is, $\lambda\iN\Mwo$ and $\lambdad\iN\Lwo$,
\resp. In addition they depend on two further parameters, the
{\em twist parameters\/} $s_1,s_2\iN\LV{\otimes_\zet^{}}\reals$.
The ordinary \g-characters are recovered when $s_1\eq0\eq s_2$ and when the
relevant lattice is the ordinary coroot lattice $\LV$ of \gb.

The ordinary characters $\chii_\lambda$ can be written as quotients of 
functions $\Xi_{\lambdab+\Rhob,k+\gv}$ and $\Xi_{\Rhob,\gv}$, which in turn 
are odd Weyl sums over Theta functions for the lattice $\LV$ \cite{KAc3}.
Similarly, the twisted characters are given by
  \be  \chii^\omego_\lambda[s_1,s_2](\tau,z) :=
  \fraC{\Xi^\omego_{\lambdab+\Rhob,k+\gv}[s_1,s_2](\tau,z)}
  {\,\Xi^\omego_{\Rhob,\gv}[s_1,s_2](\tau,z)} \,,  \Labl ch
and
  \be  \chid^\omego_{\lambdad}[s_1,s_2](\tau,z) :=
  \fraC{\Xid^\omego_{\lambdad+\Rhod,k+\gv}[s_1,s_2](\tau,z)}
  {\,\Xid^\omego_{\Rhod,\gv}[s_1,s_2](\tau,z)} \,, \Labl cd
\resp.
The functions $\Xi^\omego$ and $\Xid^\omego$ are the antisymmetrized (\wrtt
sign function $\epso$ of $\Wob$) sums 
  \be  \bearl
  \Xi^\omego_{\lambda,h}[s_1,s_2](\tau,z) := \dsty\sum_{\wb\in\Wob}
  \epso\!(\wb)\, \Theta_{\wb(\lambda),h}[s_1,s_2] (\tau,z) \,,
  \\{}\\[-.8em]
  \Xid^\omego_{\lambdad,h}[s_1,s_2](\tau,z) := \dsty\sum_{\wb\in\Wob}
  \epso\!(\wb)\, \Thetad_{\wb(\lambdad),h}[s_1,s_2] (\tau,z)
  \eear\Labl Xi
over the $\gb_\omego$-Weyl group $\Wob$, where in turn $\Theta_{\mu,h}$
and $\Thetad_{\mud,h}$ are {\em twisted Theta functions\/} associated to the 
lattices $\LVo$ and $\MVo$, \resp, at level $h$, i.e.
  \be  \bearl
  \Theta_{\lambda,h}[s_1,s_2](\tau,z) := \dsty\sum_{\beta\in\sLVo}
  \eE^{2\pi\ii\tau(\lambda+hs_1+h\beta,\lambda+hs_1+h\beta)/2h}
  \eE^{2\pi\ii(z+s_2,\lambda+hs_1+h\beta)} \,, \\{}\\[-.8em]
  \Thetad_{\lambdad,h}[s_1,s_2](\tau,z) := \dsty\sum_{\betad\in\sMVo}
  \eE^{2\pi\ii\tau(\lambdad+hs_1+h\betad,\lambdad+hs_1+h\betad)/2h}
  \eE^{2\pi\ii(z+s_2,\lambdad+hs_1+h\betad)} \,.
  \eear\Labl Th

Various properties of the functions \Erf Th and \Erf Xi are listed in
appendix \ref{a.a}. They give rise to the following properties of the 
twisted characters \Erf ch and \Erf cd.
\nxt
We can restrict our attention to a fundamental domain of the action
of $\Wo\eq\Wob\,\Ltimes\,h\LVo$ on $\Mwo$, \resp\
of $\Wob\,\Ltimes\,h\MVo$ on $\Lwo$.
Thus we need to consider only characters with 
  \be  \lambda\in\Po_k \quad{\rm and}\quad \lambdad\in\Pd_k\,,  \ee
\resp.
\nxt
{}From the result \erf{2.19} we deduce
  \be\bearl
  \chii^\omego_\lambda[s_1,s_2{+}\mu](\tau,z)
  = \eE^{2\pi\ii(\lambda,\mu)}\eE^{2\pi\ii k (s_1,\mu)}\,
  \chii^\omego_\lambda[s_1,s_2](\tau,z) \quad{\rm for}\;\
  \mu\iN\Mwo{\cap}Q^* \,,
  \\{}\\[-.8em]
  \chid^\omego_{\lambdad}[s_1,s_2{+}\mud](\tau,z)
  = \eE^{2\pi\ii(\lambdad,\mud)}\eE^{2\pi\ii k (s_1,\mud)}\,
  \chid^\omego_{\lambdad}[s_1,s_2](\tau,z )\quad{\rm for}\;\
  \mud\iN\Lwo{\cap}Q^* \,, \end{array}\labl{3.44}
where $Q^*$ is the coweight lattice of \gb.
\nxt
The behavior under the T-transformation follows from the transformation
properties \Erf78 
of the $\Xi$-functions. For the functions $\chii^\omego$ it reads
  \be  \chii^\omego_\lambda[s_1,s_2](\tau{+}1,z)
  = \eE^{-\ii\pi k (s_1,s_1)}\, T_\lambda^{}\,
  \chii^\omego_\lambda[s_1,s_1{+}s_2](\tau,z) \ee
with
  \be T_\lambda^{} := \exp\Llb 2\pi\ii [ (\lambda{+}\Rhob,
  \lambda{+}\Rhob)/2(k{+}\gv)\,{-}\,(\Rhob,\Rhob)/2\gv ] \Lrb
  = \exp(2\pi\ii (\Delta_\lambda\,{-}\,c/24)) \,.  \ee
Here it is assumed that the condition \Erf s1 is satisfied, i.e.\ that
$(s_1,\beta)\iN\zet$ for all $\beta\iN\LVo$, or in short, that
  \be  s_1 \in \Lwo \,;  \ee
this is indeed the case for all vectors $s$ that appear in table \ref{ta}, 
which will be the situation we are actually interested in.
\nxt
In the cases where $\MVo$ differs from $\LVo$, it is not an even lattice;
as a consequence the $\chid$-characters do not
transform nicely under the T-transformation. But they still do so
under T$^2$, namely
  \be  \chid^\omego_{\lambdad}[s_1,s_2](\tau{+}2,z)
  = \eE^{-2\ii\pi k(s_1,s_1)}\, (T^\omego_\lambdad)^2\,
  \chid^\omego_\lambdad[s_1,2s_1{+}s_2](\tau,z) \ee
with
  \be  T^\omego_\lambdad := \exp\Llb 2\pi\ii [ (\lambdad{+}\Rhod,
  \lambdad{+}\Rhod)/2(k{+}\gv)\,{-}\,(\Rhod,\Rhod)/2\gv ] \Lrb\,.  \Labl t2
Here again it is assumed that condition \Erf s1 is satisfied for $s_1$,
except for the case of $\gb\eq A_{2n}$, where the stronger restriction
\Erf sP applies.
\nxt
For the behavior under the S-transformation we use the formulae \Erf79 and
\erf{79t} as well as \Erf83. Introducing the matrix $\Breve S$ with entries
  \be  \Breve S_{\lambda,\mud} := |\Mwo/(k{+}\gv)\LVo|^{-1/2}_{} \,
  \ii^{(d_\omego-r_\omego)/2}\! \sum_{\wb\in \Wob}
  \epso\!(\wb)\, \eE^{-2\pi\ii (\wb(\lambda+\Rhob),\mud+\Rhod)/(k+\gv)}
  \,,  \Labl46
we obtain
  \be  \bearl
  \chii^\omego_\lambda[s_1,s_2](-\Frac1\tau,\Frac z\tau) 
  = \eE^{2\pi\ii k(s_1,s_2)} \dsty\sum_{\mud\in\Pd_k}
  \Breve S_{\lambda,\mud}\, \chid^\omego_\mud[s_2,-s_1](\tau,z) \,,
  \\{}\\[-.7em]
  \chid^\omego_\mud[s_1,s_2](-\Frac1\tau,\Frac z\tau) 
  = \eE^{2\pi\ii k(s_1,s_2)} \dsty\sum_{\lambda\in\Po_k}
  \Breve S_{\lambda,\mud}\, \chii^\omego_\lambda[s_2,-s_1] (\tau,z)
  \,.  \eear\ee
\nxt
Finally we point out that there is a close connection between the functions
$\chii^\omego_\lambda[0,0](\tau)$ and the characters $\chii'_{\mu'}$
of the twisted affine \lie\ $\gb\tw$. This is already apparent from the 
fact that the lattices $\MVo$ as listed in table \Erf tb are
precisely $2^{-1/2}$ times the lattices that appear in the Weyl group
of these twisted affine \lie s (see remark 6.7.\ of \cite{KAc3}). 
Closer inspection shows that indeed we have
 \futnot{also, level $h$ remains unchanged, and shift vectors are
related by $s_1=s'_1/\sqrt2$ and $s_2=\sqrt2 s'_2$.}
  \be  \chii'_{\mu'}(\tau,z)
  = \chid^\omego_{\mu'\!{/}\sqrt2}[0,0](2\tau{,}\sqrt2 z) \,.  \labl{chii'}
In particular, we learn that the coefficients that appear in the expansion of 
$\chid^\omego_{\mu'\!{/}\sqrt2}[0,0]$ in powers of $q$ are non-negative integers.
 \futnot{\cite[Prop.\,13.9]{KAc3}.}

\subsection{The orbifold data}

We are now in a position to compare what we have learned about the twisted 
characters $\chii^\omego_\mu[s_1,s_2]$ and $\chid^\omego_\mud[s_1,s_2]$
-- in particular their modular properties -- with the general
results about orbifolds that were presented in section 2.
This way can identify the data that characterize the WZW orbifold,
i.e.\ the functions $\chie_\mu$ and $\chiz_\mud$ and the associated
modular matrices.

First of all, the labels $\lambda$ of symmetric fields as well as
the labels $\lambdad$ in the twisted sector are precisely as defined 
above; in particular, $\lambda$ takes values in the set \Erf,;, i.e.\
is a symmetric \gb-weight that is integrable for \g\ at level $k$.
Further, let us choose the convention for the phases $\eta_\lambda$ as
introduced in \erf{utw}
so as to match the phases $\etas_\lambda$ appearing in \erf{etas}, i.e.
  \be \eta_\lambda := \etas_\lambda 
  \equiv \exp(2\pi\ii (s,\lambda))  \ee
for every symmetric weight $\lambda$.
Also, for the moment we exclude the exceptional case of outer
\auto s $\omego\eq\oc$ of $A_{2n}$, which will be treated afterwards.
Then the functions $\chie$ that 
describe the projection of symmetric fields in the untwisted sector
are given by
  \be  \chie_\lambda(2\tau) = \chii^\omego_\lambda[0,s](\tau) 
  \Labl,7
while for the twisted sector, we have to set
  \be  \chiz_\lambdad(\Frac\tau2) = \chid^\omego_\lambdad[s,0](\tau) 
  \,,  \Labl;7
with $\omego$ the diagram \auto\ and $s$ the shift vector as listed in
table \ref{ta}.
To be precise, the latter identification is unique up to possibly a phase 
that could, however, be
absorbed in the definition of $\Sz$. With the prescription given here,
it follows from our general discussion that the coefficients of an expansion
in $q$ of the expression \Erf,7 are non-negative integers.

In the exceptional $A_{2n}$ case, where $\omego\eq\oc$ and $s\eq0$,
we must in addition account for the minus
sign \erf{Zeta} that is present in the \auto\ \Erf aa of the affine \lie\
$A_{2n}\untw$. This sign amounts to a relative minus sign between
contributions from even and odd grades, hence we can immediately 
conclude that in place of \Erf,7 we now have
  \be  \chie_\lambda(2\tau) = T_\lambda^{-1/2}\,
  \chii^\oc_\lambda[0,0](\tau{+}\mbox{$\frac12$}) \,.  \Labl:7
For obtaining the analogue of \Erf;7, we need to express the
S-transformed character $\chie_\lambda(-\frac2\tau)$
as a linear combination of characters in the twisted sector, i.e.\
as a linear combination of as many power series in $q^{1/2}$ as there
are symmetric fields in the theory. When addressing this task directly 
via the expression $\chii^\oc_\lambda[0,0](-\frac1\tau{+}\frac12)$, the
calculations turn out to become clumsy.
However, the following observation can be employed to rewrite $\chii^\oc$
in a much more amenable manner. Namely, the $C_n$-modules that appear at
the even and odd grades, \resp, of the integrable highest weight modules
of the twisted affine \lie\ $A_{2n}\tw$ are distinguished by their
$C_n$ conjugacy class. It follows that a shift $\tau\Mapsto\tau+1/2$
in the argument of the $A_{2n}\tw$-characters, and hence also of the
twining characters $\chii^\oc$, can be undone by inserting the central
group element corresponding to the conjugacy class into the trace.
In the case of $A_{2n}\tw$-characters, this group element is given by 
$\exp(2\pi\ii H_{\tilde\sp})$ with $\tilde\sp$ the $C_n$-weight
$\tilde\sp\eq\Lambda_{(n)}$. Translated to the twining characters,
this becomes $\exp(2\pi\ii H_{\sp})$ with $A_{2n}$-weight\,%
 \futnote{The factor of 4 arises because the identification between
weight space and \csa\ of $C_n$ that is induced from the corresponding
identification for $A_{2n}$ differs by this factor from the standard
one.}
  \be  \sp := \Frac14\, (\Lambda_{(n)}\,{+}\,\Lambda_{(n+1)}) \,.  \Labl sp
Thus we can rewrite \Erf:7 as
  \be  \chie_\lambda(2\tau) = \chii^\oc_\lambda[0,\sp](\tau) \,.  \Labl,8
This is of the same form as the generic result \Erf,7; accordingly, in
place of \Erf;7 we now have
  \be  \chiz_\lambdad(\Frac\tau2) = \chid^\oc_\lambdad[\sp,0](\tau) 
  \,.  \Labl;8

To determine the matrices $\Se$ and $\Sz$, we calculate
  \be \bearll
  \chie_\lambda(-\Frac1\tau) \!\!
  &= \chii^\omego_\lambda[0,s](-\Frac1{2\tau})
  \\{}\\[-.8em]
  &= \dsty\sum_{\mud\in\Pd_k} \Breve S_{\lambda,\mud}\,
  \chid^\omego_\mud[s,0](2\tau)
   = \dsty\sum_{\mud\in\Pd_k} \Breve S_{\lambda,\mud}\,
  \chiz_{\mud}(\tau) \,,
  \\{}\\[-.4em]
  \chiz_\mud(-\Frac1\tau) \!\!
  &= \chid^\omego_\mud[s,0](-\Frac2\tau)
   = \dsty\sum_{\lambda\in\P_k} \Breve S_{\lambda,\mud}\,
  \chi^\omego_\lambda[0,-s](\Frac\tau2)
  \\{}\\[-.8em]
  &= \dsty\sum_{\lambda\in\P_k} \Breve S_{\lambda,\mud}\,
  \eta_\lambda^{-2}\, \chii^\omego_\lambda[0,s](\Frac\tau2)
   = \dsty\sum_{\lambda\in\P_k} \Breve S_{\lambda,\mud}\,
  \eta_\lambda^{-2}\, \chie_\lambda(\tau)
  \,, \eear \ee
while $\Tz$ is determined by employing \Erf t2 and \erf{3.44},
  \be \bearll
  \chiz_\lambdad(\tau{+}1) \!\!
  &= \chid^\omego_\lambdad[s,0](2\tau{+}2)
   = \eE^{-2\pi\ii k(s,s)}\, (T^\omego_\lambdad)^2_{}\,
  \chid^\omego_\lambdad[s,2s](2\tau)
  \\{}\\[-.8em]
  &= \eE^{-2\pi\ii k(s,s)} \eE^{2\pi\ii(\lambdad,2s)}
  \eE^{2\pi\ii k(s,2s)}\,
  (T^\omego_\lambdad)^2_{}\, \chid^\omego_\lambdad[s,0](2\tau)
  \\{}\\[-.8em]
  &= \eE^{2\pi\ii k(s,s)} \eE^{2\pi\ii(\lambdad,2s)}\,
  (T^\omego_\lambdad)^2_{}\, \chiz_\lambdad(\tau)
  \,. \eear \ee
In the case of the outer \auto\ of $A_{2n}$ we must in addition include the
appropriate shift vector $\sp$ that was introduced in \Erf sp above. It
is gratifying that according to the result \Erf sP this shift is 
precisely what is needed in order for $\chiz$ to transform nicely under
the T-operation. Furthermore, in this exceptional case we can also use 
the simple relationship $\lambdad\eq\lambda/2$.
We then find
  \be \bearl
  \Se_{\lambda,\mud} = \Breve S_{\lambda,\mud} \,,
  \\{}\\[-.55em]
  \Sz_{\mud,\lambda} = \eta_\lambda^{-2}\, \Breve S_{\lambda,\mud} \,,
  \\{}\\[-.6em]
  \Tz_\lambdad = \left\{\bearll 
  \eE^{2\pi\ii k(\sp,\sp)} \eE^{2\pi\ii(\lambda,\sp)}\, {(T_\lambda)}^{1/2}
  & {\rm for}\ \gb\eq A_{2n},\; \omego\eq\oc \,, \\[.32em]
  \eE^{2\pi\ii k(s,s)}\, \eE^{2\pi\ii(\lambdad,2s)}\,
  {(T^\omego_\lambdad)}^2 & {\rm else}\,.  \eear \right. 
  \eear \Labl,,

Let us check that the consistency conditions
$(i)$ through $(iv)$ of $\zet_2$-orbifolds
that we derived in section 2 are indeed satisfied by these matrices.
\nxt
Concerning $(i)$ we remark that the unitarity of $\Breve S$ follows by
precisely the same arguments \cite{KAc3} as for the Kac\hy Peterson 
S-matrix of affine \lie s. 
\nxt
The relation $(ii)$ between $\Se$ and $\Sz$ is manifest in \Erf,,. 
\nxt
To address property $(iii)$ we start by computing
the matrix elements $(\Breve S(\Breve S)^{\rm t}_{})_{\lambda,\lambda'}$.
This is non-zero if and only if there is an element
$\wb\iN\Wob$ such that $\wb(\lambda')\eq{-}\lambda$. This can only be
the longest element $\Breve\wb_{{\rm max}}$ of the Weyl group of 
the (horizontal) orbit \lie. Therefore we need $\lambdap\eq\lambda'$, where the
conjugation is to be taken in the orbit theory. 
The sign introduced by $\Breve w_{{\rm max}}$ cancels against
the prefactors $\ii$ for the same reason they do so in the Kac\hy Peterson
formula of the untwisted affine \lie\ that is based on the horizontal orbit
\lie. (The latter is {\em not\/} the affine orbit theory, which would be a
twisted affine \lie; here only properties of the Weyl groups of the
horizontal subalgebras matter.) Thus
  \be  \Breve S\, (\Breve S)^{\rm t}_{} = \Breve C \,,  \ee
where $\Breve C$ is the conjugation matrix of the orbit \lie. $\Breve C$ is in 
particular a permutation of order two; inspection shows that
$\Breve C$ simply coincides with the restriction of $C$ to symmetric fields,
which in turn implies that
  \be  \Ce_{\lambda,\mu} = \eta_\lambda^{-2}\, C_{\lambda,\mu}^{}
  \,.  \ee
(In fact, only when $\omega$ is inner and charge conjugation outer, then $\Breve C$
is non-trivial, while in all other cases $\Breve C$ is the identity permutation.)
Validity of property $(iii)$ thus follows from the fact that 
according to relation \Erf pm, $\eta_\lambda^2$ is a sign.
The notation for that sign in \Erf pm was chosen with hindsight; 
indeed, we simply have
  \be  \eps_\lambda^{} = \epss_\lambda  \ee
for all symmetric weights $\lambda$.
\nxt
To verify the first relation of condition $(iv)$, we insert the explicit form
\Erf46 of $\Breve S$ to deduce
 \futnot{first combine the $\eta^{-2}$ with the exponential, then replace the
summation over the alcove by a summation over the Weyl chamber, divided by the
order of the finite Weyl group; then do the lattice sum, using the fact that
$w(2s)-2s$ is an integral multiple of an element of $\LVo$ (and hence of
$\MVo$).}
  \be  \bearll  (\Sz\Se)_{\lambdad,\lambdad'} \!\!
  &= (-1)^{(d_\omego-r_\omego)/2}\, \Frac1{|\Wob|}\, \eE^{2\pi\ii(2s,\rho)}
  \\{}\\[-.8em]&\qquad
  \dsty \sum_{\wb,\wb'\in \Wob} \epso(\wb\wb')\, 
  \delta^{}_{\wb(\lambdad+\Rhod)+\wb'(\lambdad'+\Rhod+2(k+\gv)s)\,,\,0\bmod
  \!(k+\gv)\sMVo} \,.  \eear \Labl50
Now it follows immediately from the semi-direct product structure of
$\Wob$ that for every $\wb\iN\Wob$ and every $\lambdad\iN\Pd_k$ there exists a 
unique vector $\betad\iN(k{+}\gv)\MVo$ and a unique Weyl group element
$\wb'\iN\Wob$, as well as a unique $\mud\iN\Pd_k$ such that
  \be  \wb(\lambdad{+}\Rhod) = - \wb'(\mud{+}\Rhod{+}2(k{+}\gv)s{+}\betad) 
  \,. \ee
Denoting the weight $\mud$ appearing here by $\lambdadp$, it follows that
  \be  (\Sz\Se)_{\lambdad,\lambdad'} 
  = (-1)^{(d_\omego-r_\omego)/2}\, \Frac1{|\Wob|}\, \eE^{2\pi\ii(2s,\rho)}
  \sum_{\wb\in\Wob} \epso(\wb\wb')\, \delta^{}_{\lambdad,\lambdadp}
  = \pm\, \delta^{}_{\lambdad,\lambdadp} \,.  \ee
Moreover, by noticing that the square of the S-transformations acts on the
arguments as ${\rm S}^2{:}\;\tau\Mapsto\tau$, $z\Mapsto{-}z$ and considering 
characters at $z\eq0$, the sign can be seen to be $+1$.
Thus indeed $\Cz\df\Sz\Se$ is a permutation of order two.
\nxt
Finally, the second relation of $(iv)$ follows from the fact that the
element ${\rm T}{\rm S}{\rm T}^2{\rm S}{\rm T}{\rm S}{\rm T}^2{\rm S}$
of \pslz\ equals the element ${\rm S}^2$, so that
  \be  \bearll
  \dsty\sum_{\lambdad'} (\epsd P^2)_{\lambdad,\lambdad'}^{}\,
  \chid_{\lambdad'}(\tau,z) \!\!
  &= \llb {\rm T}{\rm S}{\rm T}^2{\rm S}{\rm T}{\rm S}{\rm T}^2{\rm S} \lrb
  \cdot \chid_\lambdad(\tau,z)
  \\{}\\[-1.53em]
  &= \chid_\lambdad(\tau,-z) = \dsty\sum_{\lambdad'} \Cz_{\lambdad,\lambdad'}
  \, \chid_{\lambdad'}(\tau,z)
  \,.  \eear \ee
By the linear independence of the characters, this implies that 
$\epsd P^2\eq\Cz$, as required.

The results obtained above can be checked most directly in those cases where
a different formulation of the orbifold \cft\ is available. A class of 
examples where this is the case is provided by certain conformal
embeddings \cite{scwa,babo}. One finds one infinite series, the 
embedding of $\so(n)$ at level 2 in $\su(n)$ at level 1 ($n\,{\ge}\,3$), and
one isolated case, namely $C_4$ at level 1, which is a special
subalgebra of $E_6$ at level 1. 

The simplest among these is the first member of the infinite series, i.e.\
the embedding of ${(A_1)}_4$ in ${(A_2)}_1$,
which corresponds to the D-type modular invariant of the $A_1$-\wzwt\ at
level 4. In this case, the automorphism is the diagram automorphism of
$A_2$, which also coincides with charge conjugation. There is a single
symmetric primary field, with highest weight $\vac\eq\Lambda_{(0)}
\,{\equiv}\,(0{,}0)$. Its character can be written as the sum of two 
characters of $A_1$ at level 4, $\chii_{(0{,}0)}^{}(\tau)\eq\chii_0^{\su(2)\!}
(\tau)\,{+}\,\chii_4^{\su(2)\!}(\tau)$. The orbifold chiral algebra is $A_1$ 
at level 4; thus the twining character must be the difference
of two $A_1$-characters. Indeed, one can verify explicitly the identity 
$\chie(2\tau)\eq\chii_0^{\su(2)\!}(\tau)\,{-}\,\chii_4^{\su(2)\!}(\tau)$.
Performing an S-transformation of the $A_1$-characters, we then find
$\chiz(\frac\tau2)\eq\chii_1^{\su(2)\!}(\tau)\,{+}\,\chii_3^{\su(2)\!}(\tau)$. 
We have checked explicitly this character identity, too.

Two other special situations concern inner automorphisms and charge
conjugation. They will be dealt with separately in the next two subsections.

\subsection{Inner \auto s}

In the case where the \auto\ $\omega\eq\sigmas$ is inner, many of our
results simplify. Let us describe some of the simplifications.
First note that in the inner case the rank of the
fixed point \alg\ equals the rank of \g, $\rank\gb^\sigmas\eq\rank\gb$. 
More importantly, the fact that $\omego\eq\id$ immediately implies that 
in lattice sums we just deal with the ordinary coroot lattice, and the
relevant lattice symmetries are just given by the ordinary Weyl group,
i.e.\ we have
  \be  \LVi=\MVi=\LV\,, \qquad \Lwi=\Mwi=\Lw\,, \qquad \Wib=\Wb \,.  \ee
It follows in particular that the entries of $\Tz$ now read
  \be  \Tz_\lambdad = \llb \eE^{\pi\ii k(s,s)} \eE^{2\pi\ii(\lambda,s)}\,
  T_\lambda \lrb^2_{} \,,  \Labl2t
where of course $\lambdad\,{\equiv}\,\lambda$.
Similarly, the matrix elements of $\Breve S$ coincide 
with the corresponding elements of the ordinary S-matrix of the original
\wzwt, $\Breve S_{\lambda,\mu}\eq S_{\lambda,\mu}^{}$, and indeed  
\Erf46 then is nothing but the Kac\hy Peterson formula \cite{KAc3}
for $S$. 

This result implies in particular that in the case of inner automorphisms
the number $\Nel{\lambda_1}{\lambda_2}{\lambda_3}$ that was identified
with the difference of the dimensions of the invariant subspaces under the 
action of the automorphism on the chiral blocks is in fact equal to
the fusion rules, i.e.\
$\Nel{\lambda_1}{\lambda_2}{\lambda_3}\eq\Nl{\lambda_1}{\lambda_2}{\lambda_3}$.
Indeed, using the explicit description of chiral blocks as co-invariants
(see e.g.\ \cite{tsuy,beau}), one verifies that {\em inner\/} automorphisms
of \g\ act on the chiral blocks as multiples of the identity.

The functions $\chie$ and $\chiz$ are both just shifted 
versions of ordinary characters,
  \be  \chie_\lambda(2\tau) = \chii^{}_\lambda[0,s](\tau) \,, \qquad
  \chiz_\lambdad(\Frac\tau2)= \chii^{}_\lambda[s,0](\tau) \,.  \ee
Thus the orbifold characters coming from symmetric fields read
  \be  \bearl
  \chii_{(\lambda,\psi,0)}(\tau,z)
  = \Frac12\, \llb \chii_\lambda[0,0](\tau,z) + \psi\,
  \etas_\lambda{}_{}^{-1}\, \chii_\lambda[0,s](\tau,z) \lrb \,,
  \\{}\\[-.8em]
  \chii_{(\lambda,\psi,1)}(\tau,z)
  = \Frac12\, \llb \chii_\lambda[s,0](\tau,z) +\psi\, \etas_\lambda{}_{}^{-1} 
  \eE^{-2\pi\ii k(s,s)}\, \chii_\lambda[s,s](\tau,z) \lrb\,.
  \eear \ee
Let us also mention that in the inner case it can be seen rather directly 
that the shifted characters $\chii^{}_\lambda[s,0]$ are the correct
quantities for the twisted sector.
Namely, there exists a continuous family $\{\sigma_v\}$ of
shift automorphisms of the semi-direct sum of the affine \lie\ \g\ and the
Virasoro algebra \cite{frha,levw,goom}, depending on a vector
$v$ in the \csa\ of \gb\ and acting as
  \be \bearl
  \sigma_v(H^i_n) = H^i_n + v^i K \delta_{n,0} \,, \qquad
  \sigma_v(K) = K \,, \\{}\\[-.8em]
  \sigma_v(E^\alpha_n) = E^\alpha_{n+(\alpha,v)} \,, \\{}\\[-.8em]
  \sigma_v(L_n) = L_n + (v,H_n) + \frac12(v,v) K \delta_{n,0} \,.
  \end{array}\ee
Noticing that
  \be  \chii_\lambda[s,0](\tau,z) =
  \eE^{2\pi\ii k(z,s)} \eE^{2\pi\ii\tau k(s,s)/2} \,
  \chii_\lambda[0,0](\tau,z{+}\tau s) \,,  \Labl2x
this implies that
  \be \chii_\lambda[s,0](\tau,z)
  = \Tr_{\calh_\lambda} \eE^{2\pi\ii\tau(\sigma_s(L_0))}
  \eE^{2\pi\ii(z,\sigma_s(H_0))} \,.  \ee
In words, the shifted character with shift $[s,0]$ is
precisely the character for the $\sigma_s$-twisted action of the affine \lie.
This confirms in particular once more that the coefficients in the expansion of
$\chii_\lambda[s,0](\tau,z{=}0)$ in powers of $q$ are non-negative integers.

Charge conjugation in the untwisted sector is given by
  \be  \Ce_{\lambda,\lambda'}= \etas_\lambda{}_{}^{-2}\, C_{\lambda,\lambda'}
  \,,  \Labl91
while with the help of the simple current symmetry \Erf82 of the S-matrix,
we find that the charge conjugation in the twisted sector reads
  \be  \Cz_{\lambda,\lambda'}= C_{\lambda,\J^\ssss\star\lambda'}
  \,.  \Labl92
Finally we mention that in the case of inner \auto s the matrix $P$ as defined
in formula \Erf0P differs from $S$ just by phases. Namely, using \Erf2t we have
  \be  P = \eE^{2\pi\ii k(s,s)}\, \etas TS (\etas)_{}^{-2}TST \etas
  \,.  \ee
Using $(ST)^3\eq S^2$ and again the simple current relation \Erf82,
it follows that
  \be  P_{\lambda,\lambda'}^{}
  = (\etas_\lambda\,\etas_{\lambda'})_{}^{-1}\,
  \eE^{-2\pi\ii k(s,s)}\, S_{\lambda,\lambda'}^{} \,.  \ee

\subsection{Charge conjugation}

An order-two \auto\ that is present in any arbitrary \cft, and hence is of
particular interest, is {\em charge conjugation\/}. In the WZW case this
comes from the charge conjugation \auto\ of the relevant \findim\ simple 
\lie\ \g. Data for these special \auto s are listed in the following table:
  \be \begin{array}{|l|c|c|c|c|c|}
  \hline &&&&&\\[-.82em]
  \multicolumn1{|c|} \gb & \omego & s & (s,s) & \J^\ssss & \gb^\omega
  \\[-.97em]{}&&&&&\\\hline\hline&&&&&\\[-.77em]
  A_1     & \id & \frac12\,\Lambda_{(1)} &   1/8   &  \J    & \uone        
          \nxl
  A_{2n}  & \oc & 0                      &    0    & \bfe   & B_n
          \nxl
  A_{2n+1}& \oc &\frac12\,\Lambda_{(n+1)}&(n{+}1)/8&\J^{n+1}& D_{n+1}
          \nxl
  B_{2n}  & \id & \frac12\,\Lambda_{(n)} &   n/4   &  \J^n  & B_n\oplu D_n
          \nxl
  B_{2n+1}& \id &\frac12\,\Lambda_{(n+1)}&(n{+}1)/4&\!\J^{n+1}\!& D_{n+1}\oplu B_n
          \nxl
  C_n     & \id & \frac12\,\Lambda_{(n)} &   n/8   &  \J    & A_{n-1}\oplu\uone
          \nxl
  D_{2n}  & \id & \frac12\,\Lambda_{(n)} &   n/4   & \Jv^n  & D_n\oplu D_n
          \nxl
  D_{2n+1}& \oc & \frac12\,\Lambda_{(n)} &   n/4   & \Jv^n  & B_n\oplu B_n
          \nxl
  E_6     & \oc & \frac12\,\Lambda_{(6)} &   1/2   & \bfe   & C_4
          \nxl
  E_7     & \id & \frac12\,\Lambda_{(7)} &   7/8   &  \J    & A_7
          \nxl
  E_8     & \id & \frac12\,\Lambda_{(7)} &    1    &  \bfe  & D_8
          \nxl
  F_4     & \id & \frac12\,\Lambda_{(1)} &   1/2   &  \bfe  & C_3\oplu A_1
          \nxl
  G_2     & \id & \frac12\,\Lambda_{(1)} &   1/2   &  \bfe  & A_1\oplu A_1
  \\[.47em]\hline 
  \end{array} \Labl tc

In the charge conjugation case the relevant real form of \gb\ is the compact 
real form, and we have $\Dim\gb^\omega\eq\Frac12\,(\Dim\gb -\rank\gb)$.
A quantity that we encounter in this case is the {\em \fsi\/}
of an \irrep\ of \gb, \resp\ \cite{bant5} of a primary field of the \wzwt.
By definition, this is the number $\eps_\lambda$ that takes the value
$\eps_\lambda\eq0$ for non-selfconjugate \irrep s, while in the selfconjugate
case it distinguishes between orthogonal (real) \irrep s, for which
$\eps_\lambda\eq1$, and symplectic (pseudo-real) \irrep s, which have
$\eps_\lambda\eq{-}1$. We observe that in all cases where symplectic \irrep s
occur, i.e.\ for \gb\ one of
  \be  A_{4\ell+1}\,,\quad B_{4\ell+1}\,,\quad B_{4\ell+2}\,,\quad C_r\,,\quad
  D_{4\ell+2}\,,\quad E_7\,,  \ee
the value of the \fsi\ is given by
  \be  \eps_\lambda = (-1)^{c_\lambda} \,,  \Labl36
where $c_\lambda$ is the integer
  \be  c_\lambda := 2 \cdot [ (2s,\lambdab) \bmod\zet\, ]  \,.  \ee
Inspection shows that for all \alg s \gb\ for which charge conjugation is
inner (so that all \rep s are selfconjugate),
$c_\lambda$ coincides with the conjugacy class of the
\gb-weight $\lambdab$, which in turn is twice the
monodromy charge of $\lambda$ \wrtt simple current $\J^\ssss$,
  \be  c_\lambda = 2 \, [ \Delta(\lambda)+\Delta(\J^\ssss)
  -\Delta(\J^\ssss{\star}\lambda) \bmod\zet\, ] \,.  \Labl cQ
Note that in all these cases $\J^\ssss$ has order 2, i.e.\ 
$(\J^\ssss)^2_{}\eq\bfe$, and $c_{\J^\ssss{\star}\lambda}\eq c_\lambda$.

In particular we see that the quantity $\eps_\lambda$ that was introduced in
section 2 is in this case just the exponentiated conjugacy class; it
therefore coincides with the \fsi. On the other hand, 
the general results of section 2 imply that each of the two fields 
$(\lambda,+1,0)$ and $(\lambda,-1,0)$ is self-conjugate if $\eps_\lambda\eq1$,
while they are each other's conjugate if $\eps_\lambda\eq{-}1$. This is
intuitively clear, since in the latter case the module of the original theory is
symplectic and should be split by charge conjugation into two modules
of the orbifold chiral algebra with identical Virasoro-specialized character.

\sect{Boundary conditions}

In \cite{fuSc10,fuSc11} a general prescription has been obtained by which one
can determine
the set of conformally invariant boundary conditions that preserve a
given sub\alg\ $\chir^G$ of the chiral \alg\ \chir, for the case when
$\chir^G$ is the fixed point sub\alg\ \wrt a finite abelian orbifold group $G$.
In the particular case where the orbifold group is just $G\eq\zet_2$, many
of the results of \cite{fuSc10,fuSc11} simplify enormously.
Let us present some of those results in the form in which they arise in
this specific situation.

The conformally invariant boundary conditions preserving $\chir^{\zet_2}
\,{=:}\,\chir^\omega$ are in one-to-one correspondence with the
one-dimensional irreducible \rep s of a certain \findim\ semisimple
associative commutative \alg\ \clAb, called the {\em classifying algebra\/}.
This \alg\ has a distinguished basis whose elements are
in one-to-one correspondence 
with the chiral blocks for the one-point correlation functions of bulk
fields on the disk. Let us start by describing this basis in detail
for the case of interest to us.
According to the results of \cite{fuSc10,fuSc11}, it looks as follows. Each
orbifold field $(\lambda,0,0)$
that comes from a pair of non-symmetric fields of the
original theory gives rise to two basis elements, which we label as
  \be  \tilde\Phi_\lambda \quad{\rm and}\quad \tilde\Phi_{\gs\lambda} \quad\;
  (\gs\lambda\nE\lambda) \,,  \ee
while each untwisted orbifold field $(\lambda,\psi,0)$ 
that comes from a symmetric field yields a single basis element,
which we simply denote by
  \be  \tilde\Phi_{(\lambda,\psi)} \quad{\rm with}\;\ \psi\in\{\pm1\}
  \quad\; (\gs\lambda\eq\lambda) \,.  \ee
Fields in twisted sectors of the orbifold, on the other hand, do not
correspond to any element of \clAb. 

As already mentioned,
to each element $\tilde\Phi$ in the basis of the classifying algebra
there corresponds a boundary block (i.e., a chiral block for the
one-point functions of bulk fields on the disk); we denote this 
chiral block by $\bet$. For non-symmetric fields $\lambda$,
$\bet_\lambda$ is the ordinary boundary block (also known as 
Ishibashi state) of the original \wzwt, while for symmetric $\lambda$ we
get two distinct boundary blocks $\bet_{(\lambda,\psi)}$. The `regularized
scalar products' of these boundary blocks are given by the characters 
of the orbifold theory:
  \be \bearll
  \langle\bet_\lambda |\,q^{\lolo}\,| \bet_\mu\rangle\!\!
  &= \delta_{\lambda,\mu}^{}\, \chii_{\lambda}(2\tau) \,,
  \\{}\\[-.6em]
  \langle\bet_\lambda |\,q^{\lolo}\,| \bet_{(\mu,\psi)}\rangle
  &= 0 \,,
  \\{}\\[-.6em]
  \langle\bet_{(\lambda,\psi)} |\,q^{\lolo}\,| \bet_{(\mu,\psi')}
  \rangle \!\!\!\!
  &= \delta_{\lambda,\mu}^{}\, \delta_{\psi,\psi'}^{} \cdot
  \Frac12\, ( \chii_{\lambda}(2\tau) + \psi\, \chio_\lambda(2\tau) )  
  \,. \end{array}\ee
Here $\chio_\lambda$ is the {\em twining character\/} for the \auto\
$\omega$; decomposing this \auto\ as $\omega\eq\omego\,{\circ}\,\sigmas$ 
into its diagram and inner parts (see formula \Erf om), $\chio_\lambda$
is given by the expression \erf{ch} with $s_1\eq0$ and $s_2\eq s$.

Next we list the boundary conditions that preserve the orbifold chiral 
\alg. According to \cite{fuSc10,fuSc11} they can be labelled by orbits 
\wrtt simple current
$(\vac,-1,0)$ of the orbifold, including multiplicities that take 
into account how this simple current acts by the fusion product. In the 
untwisted sector, the full orbits of this action
are $\{(\mu,1,0),(\mu,-1,0)\}$, and each such orbit gives rise to
a single \bc, which we label by $\mu$.
The fixed points are $\{(\mu,0,0)\}$; each of them provides us with two 
distinct boundary conditions, which we label by $\mu$ and $\gs\mu$. 
The twisted sector supplies us with additional \bc s. All orbits in the 
twisted sector have length two, i.e.\ they are of the form
$\{(\mud,1,1),(\mud,-1,1)\}$; accordingly
each of them amounts to a single \bc, which we label by $\mud$. 
Thus altogether the list of \bc s reads
  \be  \bearll 
  \ \, \mu & {\rm for}\ \{(\mu,1,0),(\mu,-1,0)\} \,, 
  \\{}\\[-.86em]
  \left. \bearl \mu\\{}\\[-1.09em] \gs\mu \eear\right\}
  & {\rm for}\ \{(\mu,0,0)\}
  \\{}\\[-.86em]
  \ \, \mud & {\rm for}\ \{(\mud,1,1),(\mud,-1,1)\}
  \,.  \eear \ee

The structure constants are most conveniently expressed in 
terms of a certain matrix $\tS$ which, roughly, connects the boundary
blocks to the \bc s. This matrix has two distinct types
of labels; the row index refers to boundary blocks, while the column
index refers to \bc s. In the case at hand, we obtain
  \be  \bearll
  \left.\bearl  \tS_{(\lambda,\psi),\mu} = S_{\lambda,\mu} \\{}\\[-.79em]
  \tS_{(\lambda,\psi),\mud} = \psi\eta_\lambda^{-1}\,\Se_{\lambda,\mud}
  \eear\right\}\;{\rm for}\ \gs\lambda\eq\lambda \,, \quad &
  \left.\bearl  \tS_{\lambda,\mu} = S_{\lambda,\mu} \\{}\\[-.79em]
  \tS_{\lambda,\mud} = 0 
  \eear\right\}\;{\rm for}\ \gs\lambda\nE\lambda \,.
  \eear \ee
Here $\Se$ is determined by the formulae \Erf,, and \erf{46} of section
3; it depends only on the class of the automorphism $\omega$ 
modulo inner automorphisms and reads explicitly
  \be  \Se_{\lambda,\mud} := {\cal N}_{\omego} \,
  \ii^{(d_\omego-r_\omego)/2}\! \sum_{\wb\in \Wob}
  \epso\!(\wb)\, \eE^{-2\pi\ii (\wb(\lambda+\Rhob),\mud+\Rhod)/(k+\gv)}
  \,.  \ee
Here $\Wob$ is the Weyl group of the horizontal subalgebra of the orbit 
\lie\ and $\epso$ is its sign function; $d_\omego$ and $r_\omego$ are the 
dimension and the rank of this horizontal \lie. We also recall that for
inner automorphisms the horizontal subalgebra of the orbit \lie\ 
coincides with the horizontal subalgebra \gb\ of \g; for
the outer automorphisms it is given by $C_n$ for $\gb\eq A_{2n}$, $B_{n+1}$ for 
$\gb\eq A_{2n+1}$, $C_{n-1}$ for $\gb\eq D_n$ and by $F_4$ for $\gb\eq E_6$.
Finally ${\cal N}_{\omego}$ is the real positive number which is determined
by requiring $\Se$ to be unitary.

Having obtained the matrix $\tS$,
the structure constants of the classifying algebra \clAb\ are computed
by the Verlinde-like formula
  \be \tN{\lambda_1}{\psi_1}{\lambda_2}{\psi_2}{\lambda_3}{\psi_3}
  = \sum_m \frac{ \tS_{(\lambda_1,\psi_1),m}^{}
  \tS_{(\lambda_2,\psi_2),m}^{}
  \tS_{(\lambda_3,\psi_3),m}^{-1}}{\tS_{(\vac,1),m}} \,,  \ee
where the symbol $m$ stands either for an orbit $m\eq\mu$ (with 
multiplicities) in the untwisted sector or an orbit $m\eq\mud$ 
of primary fields in the twisted sector.

We are now able to display explicitly the {\em reflection coefficients\/} 
$R^m_{(\lambda,\psi),\vac}$ and $R^m_{\lambda,\vac}$, which are the 
operator product coefficients that describe how a bulk field excites 
a boundary vacuum field $\Psi^{m,m}_\vac$ when it approaches the boundary,
according to
   \be  \phi_{l,l^+_{\phantom|}}(r\eE^{\ii\sigma}) \,\sim\,
   \sum_{l'}\sum_m
   (r^2{-}1)^{-2\Delta_l+\Delta_{l'}}_{}\, {\rm R}^m_{l,l'}\,
   \Psi^{m,m}_{l'}(\eE^{\ii\sigma}) \qquad {\rm for}\;\ r\to 1  \,. \labl{pp}
The reflection coefficients for the boundary condition $m\eq\mu$ read
  \be  {\rm R}^\mu_{(\lambda,\psi),\vac} = \frac{S_{\lambda,\mu}}
  {S_{\vac,\mu}} \,, \qquad 
  {\rm R}^\mu_{\lambda,\vac} = \frac{S_{\lambda,\mu}}
  {S_{\vac,\mu}} \,, \labl{R}
while for boundary conditions of type $m=\mud$ they are
  \be  {\rm R}^\mud_{(\lambda,\psi),\vac} = \frac{\SO_{(\lambda,\psi,0),
  (\mud,\varphi,1)}} {\SO_{(\vac,1,0),(\mud,\varphi,1)}}
  = \psi\, \eta_\lambda^{-1}\,
  \frac{\Se_{\lambda,\mud}}{\Se_{\vac,\mud}} \,, \qquad
  {\rm R}^\mud_{\lambda,\vac} = 0 \,.  \labl{Rd}

As usual the the {\em boundary state\/} $|B^m\rangle$ that is associated
to a \bc\ $m$ is a linear combination
of boundary blocks $\bet_{(\lambda,\psi)}$ and $\bet_\lambda$, with 
coefficients given by the reflection coefficients:
  \be \bearll
  |B^m\rangle \!\!\!
  &= \dsty\sum_{(\lambda,\psi)} C^m\,{\rm R}^m_{(\lambda,\psi),\vac}\,
  \bet_{(\lambda,\psi)} + \sum_\lambda C^m\, {\rm R}^m_{\lambda,\vac}\,
  \bet_\lambda
  \\{}\\[-.8em]
  &= \dsty\sum_{(\lambda,\psi)} \tS_{(\lambda,\psi),m}\,
  \bet_{(\lambda,\psi)} + \sum_\lambda \tS_{\lambda,m}\, \bet_\lambda \,.
  \end{array}\ee
Here the normalization $C^m\df\tS_{(\vac,1),m}$ ensures the correct 
normalization of the vacuum boundary field.

The boundary conditions come in two sets, corresponding to two 
automorphism types \cite{fuSc6}. We first comment on the ones labelled 
by $m\eq\mu$. In this case the reflection coefficients 
are precisely the (generalized) quantum
dimensions of the original theory. Indeed, these boundary conditions 
do not only preserve the orbifold subalgebra
$\chir^\omega$, but even the full chiral \alg\ \chir.
It is well known \cite{card9} that such \bc s are governed by the
fusion rules of the \chir-theory. In our description this behavior is
recovered as follows. The sub\alg\ \clAp\ of \clAb\ that is spanned by
the basis elements $\tilde\Phi_\lambda$ for the non-symmetric fields and 
by the sums $\tilde\Phi_{(\lambda,1)}\,{+}\,\tilde\Phi_{(\lambda,-1)}$ 
for the symmetric fields is an ideal of \clAb, and it is in fact 
isomorphic to the fusion rule \alg\ of
the WZW theory. Its \irrep s are labelled by orbits (with multiplicities)
of fields in the untwisted sector, which in turn correspond to primary
fields in the original \wzwt. Since
according to the results of \cite{fuSc10,fuSc11} the \bc s that
preserve the full bulk symmetry are precisely those that come
from the untwisted sector of the orbifold, this is not surprising at all.

The second automorphism type of boundary conditions is provided by those
which correspond to orbits in the twisted sector; they
are thus labelled with dotted indices, $m\eq\mud$. They do not preserve 
the full bulk symmetry, but only $\chir^\omega$. 
We get as many symmetry breaking boundary conditions as there
are symmetric primary fields in the original theory. 
The corresponding boundary states only involve the boundary blocks
$\bet_{(\lambda,\psi)}$ of symmetric fields; moreover, due to the 
factor of $\psi$ in \erf{Rd} only the combinations
$\bet_{(\lambda,1)}\,{-}\,\bet_{(\lambda,-1)}$ appear. The latter are just
twisted boundary blocks and reflect the breaking of the bulk symmetries.
These boundary conditions are described by the \irrep s
of a complementary ideal \clAm\ of the classifying algebra, namely the
one which is spanned by the differences
$\tilde\Phi_{(\lambda,1)}\,{-}\,\tilde\Phi_{(\lambda,-1)}$
of basis elements of \clAb.

Let us finally remark that for the determination of boundary conditions,
we only need to know the S-matrix elements that involve at least one
primary field from the untwisted sector. As a consequence, while the
determination of the WZW orbifolds becomes technically more involved
for groups $G$ other than $\zet_2$, the generalization of our results
to boundary conditions that preserve any orbifold algebra under an
arbitrary abelian group of automorphisms is straightforward.

%%%%%%%%%%%%%%%%%%%%%%%%%%%%%%%%%%%%%%%%%%%%%%%%%%%%%%%%%%%%%%%%%%%%%%%

\appendix

\sect{Appendix}\label{a.a}

Here we collect the pertinent properties of the twisted Theta functions
\Erf Th and their Weyl sums \Erf Xi that are needed in the main text.
These results are actually of a more general validity than is needed for
our present purposes.
Namely, they hold for any pair of lattices $\lvo$ and $\mvo$ of rank $r$
that satisfy the following conditions. First, $\lvo$ must be a sublattice
of $\mvo$, while $\mvo$ in turn is a sublattice of $\lvo/2$, i.e.
  \be  \lvo \subseteq \mvo \subseteq \Frac12\,\lvo \,;  \ee
this also implies that $\mwo\,{\subseteq}\,\lwo\,{\subseteq}\,\mwo/2$. 
Second, both $\lvo$ and $\sqrt2\mvo$ are even lattices; when already $\mvo$ 
is an even lattice, we put $l_\mvo\df1$, while otherwise we set $l_\mvo\df2$. 
Finally, we choose some subgroup $\wob$ of the group isometries of these 
lattices. In the case of interest to us, $\lvo\eq\LVo$ and $\mvo\eq\MVo$ are 
associated to the coroot lattice of a \findim\ simple \lie\ \gb\ in the 
manner described in subsection \ref{su.wl}, so they have $r\eq r_\omego$,
and $\wob\eq\Wob$ is the Weyl group of \gb.

In the case of inner \auto s the required properties of the 
lattices are realized trivially, since we simply have $\MVo\eq\LVo\eq\LV$,
so that in particular $l_\mvo\eq1$. For the case of outer \auto s, on the
other hand, we have $l_\mvo\eq2$, except for $\gb\eq A_{2n}$. The latter
case is in fact {\em not\/} covered by the setting described above, since
the relevant lattice $\sqrt2\MVo\eq\LVo/\sqrt2\eq Q(B_n)$ is then not 
even (see table
\Erf tb and formula \Erf37). Nevertheless this exceptional $A_{2n}$-case 
with outer \auto\ can still be treated by essentially the same methods.
Only a specific modification of the requirements to be imposed on shift
vectors is necessary, see formula \Erf sP below.

\subsection{Twisted Theta functions}

The twisted Theta functions of our interest are the lattice sums
  \be  \bearl
  \Theta_{\lambda,h}[s_1,s_2](\tau,z) := \dsty\sum_{\beta\in\lvo}
  \eE^{2\pi\ii\tau(\lambda+hs_1+h\beta,\lambda+hs_1+h\beta)/2h}
  \eE^{2\pi\ii(z+s_2,\lambda+hs_1+h\beta)} \,, \\{}\\[-.8em]
  \Thetad_{\lambdad,h}[s_1,s_2](\tau,z) := \dsty\sum_{\betad\in\mvo}
  \eE^{2\pi\ii\tau(\lambdad+hs_1+h\betad,\lambdad+hs_1+h\betad)/2h}
  \eE^{2\pi\ii(z+s_2,\lambdad+hs_1+h\betad)} \,,
  \eear\Labl tH
where $\tau$ is in the upper complex half plane,
$z,s_1,s_2\iN\lvo{\otimes_\zet^{}}\reals$, $h\iN\zetplus$, and
$\lambda\iN\mwo$, $\lambdad\iN\lwo$.
Some properties of these functions are the following.
\nxt
$\Theta_{\lambda,h}[s_1,s_2]$ depends on $\lambda$ only modulo $h\lvo$. Also,
it depends on $\lambda$ and $s_1$ only via the combination $\lambda\,{+}\,hs_1$;
but still it proves to be convenient to keep both parameters.
\nxt
For any automorphism (i.e., lattice preserving isometry) $w$ of
the lattice $\lvo$ one has
  \be \Theta_{\lambda,h}[w(s_1),w(s_2)](\tau,z) =
  \Theta_{w^{-1}(\lambda),h}[s_1,s_2](\tau,w^{-1}(z)) \,;  \labl{4.4}
in particular, the two functions coincide as functions of $\tau$.
(When $\lvo$ is the coroot lattice or
root lattice of a \findim\ simple \lie\ \gb, every \auto\ 
is an element of the product of the Weyl group $\Wb$ of \gb\ with
certain outer automorphisms.)
\nxt
Manifestly, $s_1$ is defined only modulo $\lvo$ and $\mvo$, \resp.
\nxt
Concerning shifts in $s_2$, we have
  \be  \Theta_{\lambda,h}[s_1,s_2{+}\mu](\tau,z) =
  \eE^{2\pi\ii (h(\mu,s_1)+(\mu,\lambda))}\,
  \Theta_{\lambda,h}[s_1,s_2](\tau,z)  \labl{1.3}
for every $\mu\iN\mwo$. For $\mu\eq\beta\iN\lvo\,{\supseteq}\,\lwo$
this reduces to
  \be  \Theta_{\lambda,h}[s_1,s_2{+}\beta](\tau,z) =
  \eE^{2\pi\ii h (\beta,s_1)}\, \Theta_{\lambda,h}[s_1,s_2](\tau,z) \,.
  \ee
\nxt
Next we study the T-transformation $T{:}\;\tau\Mapsto\tau+1$,
$z\Mapsto z$. Using the fact that $(\beta,\beta)\iN2\zet$ and
$(\lambda,\beta)\iN\zet$ for all $\beta\iN\lvo$ and
all $\lambda\iN\mwo\,{\subseteq}\,\lwo$, we see that the twisted Theta 
functions get multiplied by a phase,
  \be  \Theta_{\lambda,h}[s_1,s_2](\tau{+}1,z) =
  \eE^{2\pi\ii (\frac{(\lambda,\lambda)}{2h} - \frac h2\,(s_1,s_1))}\,
  \Theta_{\lambda,h}[s_1,s_1{+}s_2](\tau,z) \,,  \Labl,6
provided that 
  \be  (s_1,\beta)\in\zet \quad{\rm for\ all}\;\ \beta\iN\lvo \,.  \Labl s1
This nicely
reflects the fact that the T-transformation is the element of the mapping
class group of the torus that adds an $a$-cycle to the $b$-cycle of the torus.
\nxt
All these properties apply analogously to the twisted Theta functions
$\Thetad_{\lambdad,h}[s_1,s_2]$ that are defined with the lattice $\mvo$ in
place of $\lvo$. However, in the case of the T-transformation we now have
to take the $l_\mvo$th power in order for the twisted Theta function to
acquire just a phase. More precisely, under the condition that
  \be  2\,(\lambdad,\betad) \in \zet  \Labl-1
as well as
  \be  (\betad,\betad) + 2\,(s_1,\betad) \in \zet  \Labl-2
for all $\betad\iN\mvo$ and all $\lambdad\iN\lwo$, it follows that
  \be  \Thetad_{\lambdad,h}[s_1,s_2](\tau{+}l_\mvo,z) =
  \eE^{2\pi\ii l_\mvo (\frac{(\lambdad,\lambdad)}{2h} - \frac h2\,(s_1,s_1))}\,
  \Thetad_{\lambdad,h}[s_1,l_\mvo s_1{+}s_2](\tau,z) \,.  \Labl;6
The relation \Erf-1 is indeed satisfied, since $2\betad\iN\lvo\,{\equiv}\,
{(\lwo)}^*$, while \Erf-2 reduces to $2(s_1,\betad)\iN\zet$ because the
lattice $\sqrt2\mvo$ is even, which in turn is valid whenever \Erf s1 holds,
again as a consequence of $2\betad\iN\lvo$.
\nxt
In the previous reasoning we have used explicitly the requirement that
$\sqrt2\lvo$ is an even lattice. As already noted above, 
this is not fulfilled in the case where $\gb\eq A_{2n}$ and $\omego\eq\oc$.
Now the relevant lattice $\lvo$ is in this special case spanned by the 
elements $\alphad^{(i)}\eq\alpha^{(i)}{+}\alpha^{(2n+1-i)}$, $i\eq1,2,...
\,,n$, of the $A_{2n}$ root lattice. Inspection shows that
the condition \Erf-1 is then still satisfied, for the same
reason as before, but condition \Erf-2 no longer reduces to \Erf s1. Rather,
one must distinguish between the cases where $(\betad,\betad)$ is integral
and those where it lies in $\zet{+}1/2$. In the former case the coefficient
of $\alphad^{(n)}$ in an expansion of $\betad$ \wrtt $\alphad^{(i)}$ is
even, while in the latter case it is odd. It follows that the requirement
on $s_1$ is that
  \be  s_1 = \sp + \Frac12\, \sum_{i=1}^n m_i\, (\Lambda_{(i)}\,{+}\,
  \Lambda_{(2n+1-i)})  \Labl sP
with $m_i\iN\zet$ for $i\eq1,2,...\,,n$ and
  \be  \sp = \Frac14\, (\Lambda_{(n)}\,{+}\,\Lambda_{(n+1)}) \,,  \ee
which is the $A_{2n}$-weight already encountered in \Erf sp.
\nxt
To study also the S-transformation $S{:}\;\tau\Mapsto{-}1/\tau$,
$z\Mapsto z/\tau$, we consider both types of functions together.
Poisson resummation shows that twisted Theta functions of one type
transform into linear combinations of twisted Theta functions of the
other type, according to
  \be \bearll \Theta_{\lambda,h}[s_1,s_2](-\Frac1\tau,\Frac z\tau) \!\!
  &= |\lwo/\lvo|^{-1/2} \, h^{-r/2} (-\ii\tau)^{r/2}\, \eE^{\ii\pi h(z,z)/\tau}
  \\{}\\[-.7em]&\quad
  \eE^{2\pi\ii h(s_1,s_2)} \dsty\sum_{\mud\in\lwo/h\mvo}
  \eE^{-2\pi\ii(\mud,\lambda)/h}\, \Thetad_{\mud,h}[s_2{,}{-}s_1](\tau,z)
  \\{}\\[-.7em]
  \Thetad_{\lambdad,h}[s_1,s_2](-\Frac1\tau,\Frac z\tau) \!\!
  &= |\mwo/\mvo|^{-1/2} \, h^{-r/2} (-\ii\tau)^{r/2}\, \eE^{\ii\pi h(z,z)/\tau}
  \\{}\\[-.7em]&\quad
  \eE^{2\pi\ii h(s_1,s_2)} \dsty\sum_{\mu\in\mwo/h\lvo}
  \eE^{-2\pi\ii(\mu,\lambdad)/h}\, \Theta_{\mu,h}[s_2{,}{-}s_1](\tau,z)
  \,. \eear \Labl39
The behavior of the shift parameters is again in accordance with
the action of the mapping class group element on the fundamental cycles.
Notice that the factor $\exp(-2\pi\ii(\mud,\lambda)/h)$ is single-valued
when $\mud$ is defined modulo $h\mvo$, because $\lambda\iN\mwo$, and also when
$\lambda$ is defined modulo $h\lvo$, because $\mud\iN\lwo$; then the 
isomorphisms $\mwo/\lvo\,{\cong}\,(\lwo/\mvo)^*_{}\,{\cong}\,\lwo/\mvo$ 
(compare \erf{1.5})
imply that there are as many $\Theta$-functions as $\Thetad$-functions.
\nxt
The matrices\,%
 \futnote{Just like for the ordinary characters, the further factor
$\eE^{\ii\pi h(z,z)/\tau}$ gets absorbed by the inhomogeneous transformation 
$u\,{\mapsto}\,u-(z,z)/2\tau$ of a third variable $u$ on which these functions 
should depend. For brevity that variable is suppressed throughout this paper.}
  \be \bearl
  \cals\Untw_{\lambda,\mud} = |\lwo/\lvo|^{-1/2} h^{-r/2}\,
  \eE^{2\pi\ii h(s_1,s_2)}\, \eE^{-2\pi\ii(\mud,\lambda)/h} \,,
  \\{}\\[-.7em]
  \cals\Tw_{\lambdad,\mu} = |\mwo/\mvo|^{-1/2} h^{-r/2}\,
  \eE^{2\pi\ii h(s_1,s_2)}\, \eE^{-2\pi\ii(\mu,\lambdad)/h}
  \end{array}\ee
that arise in the S-transformation \Erf39 are not unitary; rather,
  \be \bearll
  \dsty\sum_{\mud\in\lwo/h\mvo} \cals\Untw_{\lambda,\mud} (\cals\Untw
  _{\lambda',\mud})^*_{} = |\mvo/\lvo|^{-1}\, \delta_{\lambda,\lambda'} \,,
  \ & \dsty\sum_{\lambda\in\mwo/h\lvo} \cals\Untw_{\lambda,\mud}
  (\cals\Untw_{\lambda,\mud'})^*_{} = |\lwo/\mwo|^{-1}\, \delta_{\mud,\mud'} \,,
  \\{}\\[-.7em]
  \dsty\sum_{\mud\in\lwo/h\mvo} \cals\Tw_{\mud,\lambda}
  (\cals\Tw_{\mud,\lambda'})^*_{} = |\lwo/\mwo|\, \delta_{\lambda,\lambda'} \,,
  & \dsty\sum_{\lambda\in\mwo/h\lvo} \cals\Tw_{\mud,\lambda}
  (\cals\Tw_{\mud',\lambda})^*_{} = |\mvo/\lvo|\, \delta_{\mud,\mud'} \,.
  \end{array}\labl{orth}
Because of $|\lwo/\mwo|\eq|\mvo/\lvo|$ (compare formula \erf{1.6}),
both for $\cals\Untw$ and $\cals\Tw$ the two relations come with
the same factor as they should.

\subsection{Twisted $\Xi$-functions}

Using the $\wob$-character $\epsv$ we now antisymmetrize
the $\Theta$- and $\Thetad$-functions according to
  \be  \bearl
  \Xi^\omegv_{\lambda,h}[s_1,s_2](\tau,z) := \dsty\sum_{w\in\wob}
  \epsv(w)\, \Theta_{\wb(\lambda),h}[s_1,s_2] (\tau,z) \,,
  \\{}\\[-.8em]
  \Xid^\omegv_{\lambdad,h}[s_1,s_2](\tau,z) := \dsty\sum_{w\in\wob}
  \epsv(w)\, \Thetad_{\wb(\lambdad),h}[s_1,s_2] (\tau,z) \,.
  \eear \ee
These functions inherit most of their
properties from those of the twisted Theta functions. Let us list some of them.
\nxt
They are invariant under shifts by the lattice $h\lvo$ and $h\mvo$, \resp:
  \be \bearl
  \Xi^\omegv_{\lambda+h\beta}[s_1,s_2] = \Xi^\omegv_\lambda[s_1,s_2]
  \quad\mbox{for all }\,\beta\iN\lvo \,,  \\{}\\[-.6em]
  \Xid^\omegv_{\lambdad+h\betad}[s_1,s_2] = \Xid^\omegv_{\lambdad}[s_1,s_2]
  \quad\mbox{for all }\,\betad\iN\mvo \,.  \eear \ee
\nxt
They depend only on the combination $\lambda\,{+}\,hs_1$ and
$\lambdad\,{+}\,hs_1$, \resp. Thus as a parameter of $\Xi^\omegv$,
$s_1$ can be regarded as being defined modulo $\lvo$, and as a parameter of
$\Xid^\omegv$, modulo $\mvo$; still, in both cases we will keep both arguments.
\nxt
Because of \erf{1.3} we have
  \be\bearl
  \Xi^\omegv_{\lambda,h}[s_1,s_2{+}\mu](\tau,z)
  = \eE^{2\pi\ii h(\mu,s_1)} \dsty\sum_{w\in\wob}
  \epsv(w)\, \eE^{2\pi\ii(\mu,w(\lambda))}\,
  \Theta_{\wb(\lambda),h}[s_1,s_2] (\tau,z) \,,
  \\{}\\[-.7em]
  \Xid^\omegv_{\lambdad,h}[s_1,s_2{+}\mud](\tau,z)
  = \eE^{2\pi\ii h(\mud,s_1)} \dsty\sum_{w\in\wob}
  \epsv(w)\, \eE^{2\pi\ii(\mud,w(\lambdad))}\,
  \Thetad_{\wb(\lambdad),h}[s_1,s_2] (\tau,z)
  \end{array}\labl{2.18}
for all $\mu\iN\mwo$ and $\mud\iN\lwo$.
\nxt 
At $z\eq0$ those $\Xi^\omegv$-functions for which $s_2$ lies in $\lvo$ have 
integral coefficients in the expansion in powers of $q\eq\exp(2\pi\ii\tau)$.
\nxt
As the lattices $\lvo$ an $\mvo$ are mapped to themselves under $\wob$, the 
$\Xi^\omegv$- and $\Xid^\omegv$-functions are $\wob$-odd (that is, $\epsv$-twisted):
  \be \bearl
  \Xi^\omegv_{w(\lambda)}[s_1,s_2] = \epsv(w)\, \Xi^\omegv_\lambda[s_1,s_2]
  \quad\mbox{for all }w\iN\wob\,, \\{}\\[-.7em]
  \Xid^\omegv_{w(\lambdad)}[s_1,s_2] = \epsv(w)\, \Xid^\omegv_\lambdad[s_1,s_2]
  \quad\mbox{for all }w\iN\wob\,. \eear \ee
\nxt
Owing to \Erf,6, \Erf;6 and \erf{2.18}, \resp, under the above-mentioned 
restrictions on the shift vector $s_1$ we have
  \be \bearll
  \Xi^\omegv_{\lambda,h}[s_1,s_2](\tau{+}1,z) \!\!\!
  &= \eE^{2\pi\ii (\frac{(\lambda,\lambda)}{2h} - \frac h2\,(s_1,s_1))}\,
  \Xi^\omegv_{\lambda,h}[s_1,s_1{+}s_2](\tau,z) \,,
  \\{}\\[-.8em]
  \Xid_{\lambdad,h}[s_1,s_2](\tau{+}l_\mvo,z) \!\!\!
  &= \eE^{2\pi\ii l_\mvo (\frac{(\lambdad,\lambdad)}{2h}-\frac h2\,(s_1,s_1))}\,
  \Xid^\omegv_{\lambdab,h}[s_1,l_\mvo s_1{+}s_2](\tau,z)  \eear \Labl78
for the behavior under the T-operation.

\subsection{The S-transformation}

The transformation of $\Xi^\omegv$ and $\Xid^\omegv$ under the S-operation
could still be discussed for the general situation studied so far, provided
that appropriate properties of the group $\wob$ are imposed. For brevity we
now restrict, however, our attention to the specific case that is of interest
in the main text. Thus in particular $\wob\eq\Wob$ is the Weyl group of \gb, 
while $\epsv\eq\epso$ is the sign function of the Coxeter group $\Wob$.
Combining the result \Erf39 with the fact that the group $\Wob$ acts on 
all four lattices by isometries and with the \rep\ property of $\epsv$, 
it follows that in this case we have
  \be  \bearll  \Xi^\omego_{\lambda,h}[s_1,s_2](-\Frac1\tau,\Frac z\tau) \!\!
  &= |\Lwo/\LVo|^{-1/2} h^{-r/2}(-\ii\tau)^{r/2}\, \eE^{\ii\pi h(z,z)/\tau}\,
  \eE^{2\pi\ii h(s_1,s_2)}
  \\{}\\[-.8em]&\quad
  \dsty\sum_{\mud\in\PPd_h} \sum_{\wb\in\Wob} \epso(\wb)\,
  \eE^{-2\pi\ii(\mud,\wb(\lambda))/h}\, \Xid^\omego_{\mud,h}[s_2,-s_1](\tau,z)
  \eear \Labl79
and
  \be  \bearll  \Xid^\omego_{\lambdad,h}[s_1,s_2](-\Frac1\tau,\Frac z\tau) \!\!
  &= |\Mwo/\MVo|^{-1/2} h^{-r/2}(-\ii\tau)^{r/2}\, \eE^{\ii\pi h(z,z)/\tau}\,
  \eE^{2\pi\ii h(s_1,s_2)}
  \\{}\\[-.8em]&\quad
  \dsty\sum_{\mu\in\PPo_h} \sum_{\wb\in\Wob} \epso(\wb)\,
  \eE^{-2\pi\ii(\mu,\wb(\lambdad))/h}\, \Xi^\omego_{\mu,h}[s_2,-s_1](\tau,z) \,,
  \eear \labl{79t}
\resp. Moreover,
we have $\PPo_\gv\eq\{\Rhob\}$ and $\PPd_\gv\eq\{\Rhod\}$, so that
according to \Erf79 and \erf{79t} we find
  \be
  \bearll  \Xi^\omego_{\Rhob,\gv}[s_1,s_2](-\Frac1\tau,\Frac z\tau) \!\!
  &= |\Lwo/\LVo|^{-1/2} (\gv)^{-r_\omego/2} (-\ii\tau)^{r_\omego/2}
  \eE^{\ii\pi\gv(z,z)/\tau}
  \\{}\\[-.8em]&\quad
  \eE^{2\pi\ii\gv(s_1,s_2)}\,\facrr\, \Xid^\omego_{\Rhod,\gv}[s_2,-s_1](\tau,z) \,,
  \\{}\\[-.5em]
  \Xid^\omego_{\Rhod,\gv}[s_1,s_2](-\Frac1\tau,\Frac z\tau) \!\!
  &= |\Mwo/\MVo|^{-1/2} (\gv)^{-r_\omego/2} (-\ii\tau)^{r_\omego/2}
  \eE^{\ii\pi\gv(z,z)/\tau}
  \\{}\\[-.8em]&\quad
  \eE^{2\pi\ii\gv(s_1,s_2)}\,\facrr\, \Xi^\omego_{\Rhob,\gv}[s_2,-s_1](\tau,z) \,,
  \eear \Labl80
where
  \be  \facrr := \sum_{\wb\in\Wob} \epso\!(\wb)\, \eE^{-2\pi\ii(\Rhod,
  \wb(\Rhob))/\gv} \,. \ee
The absolute value of the number $\facrr$ follows from the general 
orthogonality relations \erf{orth}:
  \be  |\facrr| = |\Lwo/\gv\MVo|^{-1/2}_{} \equiv |\Mwo/\gv\LVo|^{-1/2}_{} 
  \,.  \ee
To determine its phase, we employ the 
denominator identity of the \findim\ orbit \lie\ to find
  \be  \facrr = \prod_{\alphad\in\Delta_\omego} \Llb {-}2\ii
  \sin(\Frac{\pi(\Rhod,\alphad)}{\gv}) \Lrb  \ee
with $\Delta_\omego$ the set of symmetric \gb-roots,
which implies that $\facrr{\cdot}\,\ii^{|\Delta_\omego|}$
is a positive real number. Collecting these results, we can conclude that
  \be  \bearll
  \Xi^\omego_{\Rhob,\gv}[s_1,s_2](-\Frac1\tau,\Frac z\tau) \!\!
  &= |\MVo/\LVo|^{-1/2}_{} (-\ii)^{d_\omego/2} \tau^{r_\omego/2}
  \eE^{\ii\pi\gv(z,z)/\tau}\,
  \eE^{2\pi\ii\gv(s_1,s_2)}\, \Xid^\omego_{\Rhod,\gv}[s_2,-s_1](\tau,z)\,,
  \\{}\\[-.4em]
  \Xid^\omego_{\Rhod,\gv}[s_1,s_2](-\Frac1\tau,\Frac z\tau) \!\!
  &= |\MVo/\LVo|^{1/2}_{} (-\ii)^{d_\omego/2} \tau^{r_\omego/2}
  \eE^{\ii\pi\gv(z,z)/\tau}
  \\{}\\[-.8em]&\quad
  \eE^{2\pi\ii\gv(s_1,s_2)}\, \Xi^\omego_{\Rhob,\gv}[s_2,-s_1](\tau,z) \,,
  \eear \Labl83
where $d_\omego$ is the dimension of the \findim\ orbit \lie.

Finally we consider the relation \erf{2.18} for the special case where
the vectors $\mu\iN\mwo$ and $\mud\iN\lwo$ lie in addition in the 
{\em co\/}weight lattice of \gb. In this case we have 
$(\mu,\wb(\lambda))\eq(\mu,\lambda)\bmod\zet$ and
$(\mud,\wb(\lambda))\eq(\mud,\lambda)\bmod\zet$ 
for all $\wb\iN\Wb$, and hence
  \be\bearl
  \Xi^\omegv_{\lambda,h}[s_1,s_2{+}\mu](\tau,z)
  = \eE^{2\pi\ii(h(\mu,s_1)+(\mu,\lambda))}\,
  \Xi^\omegv_{\lambda,h}[s_1,s_2](\tau,z) \,,
  \\{}\\[-.7em]
  \Xid^\omegv_{\lambdad,h}[s_1,s_2{+}\mud](\tau,z)
  = \eE^{2\pi\ii(h(\mud,s_1)+(\mud,\lambdad))}\,
  \Xid^\omegv_{\lambdad,h}[s_1,s_2](\tau,z) \,.
  \end{array}\labl{2.19}

\vskip2em
\noindent
\small
{\bf Acknowledgement}:\\ C.S.\ would like to thank UC Berkeley for 
hospitality while part of this work was done.

%%%%%%%%%%%%%%%%%%%%%%%%%%%%%%%%%%%%%%%%%%%%%%%%%%%%%%%%%%%%%%%%%%%%%%%

\newpage
%\vskip3em
\small
 \renewcommand\wb{\,\linebreak[0]} \def\wB {$\,$\wb}
 \newcommand\Bi[1]    {\bibitem{#1}}
 \renewcommand\J[5]   {{\sl #5}, {#1} {#2} ({#3}) {#4} }
 \newcommand\PhD[2]   {{\sl #2}, Ph.D.\ thesis (#1)}
 \newcommand\Prep[2]  {{\sl #2}, preprint {#1}}
 \newcommand\BOOK[4]  {{\em #1\/} ({#2}, {#3} {#4})}
 \newcommand\inBO[7]  {{\sl #7}, 
                      in:\ {\em #1}, {#2}\ ({#3}, {#4} {#5}), p.\ {#6}}
 \newcommand\Erra[3]  {\,[{\em ibid.}\ {#1} ({#2}) {#3}, {\em Erratum}]}
 \def\jf    {J.\ Fuchs}
 \def\anop  {Ann.\wb Phys.}
 \def\aspm  {Adv.\wb Stu\-dies\wB in\wB Pure\wB Math.}
 \def\comp  {Com\-mun.\wb Math.\wb Phys.}
 \def\duki  {Duke\wB Math.\wb J.\ (Int.\wb Math.\wb Res.\wb Notes)}
 \def\foph  {Fortschr.\wb Phys.}
 \def\ijmp  {Int.\wb J.\wb Mod.\wb Phys.\ A}
 \def\jams  {J.\wb Amer.\wb Math.\wb Soc.}
 \def\joal  {J.\wB Al\-ge\-bra}
 \def\jomp  {J.\wb Math.\wb Phys.}
 \def\lmslns{London Math.\ Soc.\ Lecture Note Series \# }
 \def\maan  {Math.\wb Annal.}
 \def\mpla  {Mod.\wb Phys.\wb Lett.\ A}
 \def\nuci  {Nuovo\wB Cim.}
 \def\nupb  {Nucl.\wb Phys.\ B}
 \def\pams  {Proc.\wb Amer.\wb Math.\wb Soc.}
 \def\phlb  {Phys.\wb Lett.\ B}
 \def\phrd  {Phys.\wb Rev.\ D}
 \def\phrl  {Phys.\wb Rev.\wb Lett.}
 \def\AP     {{Academic Press}}
 \def\CUP    {{Cambridge University Press}}
 \def\NH     {{North Holland Publishing Company}}
 \def\SV     {{Sprin\-ger Ver\-lag}}
 \def\Ad     {{Amsterdam}}
 \def\Be     {{Berlin}}
 \def\Ca     {{Cambridge}}
 \def\NY     {{New York}}

\end{document}